\documentclass[aps,prd,twocolumn,superscriptaddress,nofootinbib]{revtex4-2}

\usepackage{array}
\usepackage[colorlinks=true,citecolor=red,urlcolor=blue]{hyperref}

\usepackage{amssymb}
\usepackage{amsmath,color}
\usepackage{booktabs}
\usepackage{graphicx}
\usepackage{bbm}
\usepackage{verbatim}
\usepackage[T1]{fontenc}
\usepackage[utf8]{inputenc}
\usepackage[normalem]{ulem}

\usepackage{url}
\usepackage{xcolor}
\usepackage{caption}
\usepackage{subcaption}
\usepackage{orcidlink}

\usepackage{diagbox}
\usepackage{color}
\usepackage{multirow}
\usepackage{hhline}
\usepackage{tabularx}
\usepackage{placeins}

\providecommand{\Eprint}[1]{}
\newcommand{\orcid}[1]{\href{https://orcid.org/#1}{\hspace{0.5mm}\raisebox{-0.5ex}{\includegraphics[height=2.0ex]{orcid.png}}}}

\begin{document}


\title{Parameter estimation of eccentric massive black hole binaries with LISA and its cosmological implications}


\author{Jia-Hao Zhong\orcidlink{0009-0001-3101-3774}}
\affiliation{School of Physics and Technology, Wuhan University, Wuhan 430072, China} 
 
\author{Jin-Zhao Yang\orcidlink{0000-0002-4826-6014}}
\affiliation{School of Physics and Technology, Wuhan University, Wuhan 430072, China} 

\author{Tao Yang\orcidlink{0000-0002-2161-0495}}
\email{Corresponding author: yangtao@whu.edu.cn}
\affiliation{School of Physics and Technology, Wuhan University, Wuhan 430072, China}

\author{Xu-Heng Ding\orcidlink{0000-0001-8917-2148}}
\affiliation{School of Physics and Technology, Wuhan University, Wuhan 430072, China}

\author{Xi-Long Fan\orcidlink{0000-0002-8174-0128}}
\affiliation{School of Physics and Technology, Wuhan University, Wuhan 430072, China}

\author{Kai Liao\orcidlink{0000-0002-4359-5994}}
\affiliation{School of Physics and Technology, Wuhan University, Wuhan 430072, China}

\author{Bei You\orcidlink{0000-0002-8231-063X}}
\affiliation{School of Physics and Technology, Wuhan University, Wuhan 430072, China}

\date{\today}

\begin{abstract}
Future space-based gravitational wave (GW) observatories such as LISA will detect massive black hole binaries (MBHBs), which are expected to be accompanied by electromagnetic counterparts, thereby providing bright standard sirens for cosmology. The orbital eccentricity of MBHBs can significantly improve the parameter estimation of GWs because the multiple harmonics induced by eccentricity provide additional information and help break down the degeneracies among waveform parameters. In this paper, we use the EccentricFD waveform and construct 5-year GW event catalogs for LISA under three population models (PopIII, Q3d, and Q3nod). For the three models, we find that an initial eccentricity of $e_0=0.4$ at $10^{-4}$ Hz yields improvements in sky localization and distance inference by a factor of $\mathcal{O}(10)$ in the best cases. As a consequence, the average number of bright siren candidates increases substantially: from 8 to 11 (PopIII), 6 to 12 (Q3d) and 13 to 24 (Q3nod). This increase in event number, together with enhanced localization and distance inference, leads to tighter cosmological constraints. In the $\Lambda$CDM model, for instance, the relative uncertainty on $H_0$ is reduced from $8.17\%$ to $4.35\%$ for the Q3d model, corresponding to an improvement of approximately $47\%$. We also investigate the improvement in constraints on the dark energy equation of state and modified GW propagation when combining bright sirens with the latest cosmic microwave background data. These results demonstrate that eccentricity is a remarkably significant feature in GW detection and parameter estimation, enabling more accurate measurements of the Universe with future space-based observatories.

\end{abstract}

\maketitle

\section{Introduction}
The precise measurement of cosmic microwave background (CMB) anisotropies has initiated the era of precision cosmology~\cite{WMAP:2003elm,WMAP:2003ivt}. However, these observations have revealed a nearly $5\sigma$ tension between the value of the Hubble constant inferred from the $Planck$ CMB, $H_0 \sim 67~\mathrm{km\ s^{-1}\ Mpc^{-1}}$~\cite{Planck:2018vyg}, and that measured by the SH0ES project in the late Universe, $H_0 \sim 72~\mathrm{km\ s^{-1}\ Mpc^{-1}}$~\cite{Riess:2021jrx}. This so-called Hubble tension has become one of the most significant and persistent challenges in modern cosmology [see Refs.~\cite{DiValentino:2021izs,Abdalla:2022yfr,Perivolaropoulos:2021jda,CosmoVerseNetwork:2025alb} and references therein for recent reviews]. Thus, an independent way to measure the Hubble constant is significantly important and needed.

The gravitational wave (GW) standard siren method, first proposed in~\cite{Schutz:1986gp}, provides one of the most promising and independent approaches to cosmological inference and has been extensively studied in the literature~\cite{Dalal:2006qt,Holz:2005df,Nissanke:2009kt,LIGOScientific:2017adf,Chen:2017rfc,Borhanian:2020vyr,DelPozzo:2011vcw,Nair:2018ign,LIGOScientific:2018gmd,DES:2019ccw,Gray:2019ksv,Finke:2021aom,LIGOScientific:2021aug,LIGOScientific:2025jau,Chernoff:1993th,Markovic:1993cr,Taylor:2012db,Farr:2019twy,Mastrogiovanni:2021wsd,Ezquiaga:2020tns,Ezquiaga:2022zkx,LIGOScientific:2025jau,Oguri:2016dgk,Mukherjee:2019wcg,Mukherjee:2020hyn,Bera:2020jhx,Mukherjee:2022afz,Yang:2021qge,Tamanini:2016zlh,LISACosmologyWorkingGroup:2019mwx,Mangiagli:2023ize,Zhan:2025jqg}. Analogous to the use of standard candles in astronomy, standard sirens exploit the distance–redshift relation to estimate the Hubble constant. Beyond constraining the expansion history of the Universe, it also enables precision tests of general relativity and its possible extensions on cosmological scales~\cite{LISACosmologyWorkingGroup:2019mwx,LIGOScientific:2025jau}. The luminosity distance of a GW source can be directly inferred from its waveform. Unlike standard candles, standard sirens do not rely on astronomical distance ladders or calibration. However, GW observations alone do not provide the source redshift, due to the intrinsic degeneracy between mass and redshift in the waveform. Additional information is therefore required to obtain the source redshift. If the redshift can be directly obtained from electromagnetic (EM) counterparts, a scenario known as ``bright siren''~\cite{Dalal:2006qt,Holz:2005df,Nissanke:2009kt}, then the distance–redshift relation can be used to probe the expansion history of the Universe. To date, however, ground-based GW detectors have yielded only a single confirmed bright siren, the binary neutron star merger GW170817~\cite{LIGOScientific:2017vwq,LIGOScientific:2017ync,LIGOScientific:2017zic}, which constrained the Hubble constant with a precision of $\sim14\%$~\cite{LIGOScientific:2017adf}. Although future detections are expected to improve this precision, forecasts indicate that the number of bright sirens accessible to LIGO will remain limited~\cite{Yang:2021qge}. For GW sources without detected EM counterparts, commonly referred to as ``dark sirens'', additional techniques are required to statistically infer the source redshift. A variety of methods have been proposed including using astrophysically motivated source-frame mass distributions~\cite{Chernoff:1993th,Markovic:1993cr,Taylor:2012db,Farr:2019twy,Mastrogiovanni:2021wsd,Ezquiaga:2020tns,Ezquiaga:2022zkx}, obtaining the statistical redshift information from the galaxy catalogs~\cite{Schutz:1986gp,Borhanian:2020vyr,DelPozzo:2011vcw,Nair:2018ign,LIGOScientific:2018gmd,DES:2019ccw,Gray:2019ksv,Finke:2021aom,LIGOScientific:2021aug,LIGOScientific:2025jau}, cross-correlating with large-scale structure tracers~\cite{Oguri:2016dgk,Mukherjee:2019wcg,Mukherjee:2020hyn,Bera:2020jhx,Mukherjee:2022afz}, or exploiting tidal effects in BNS mergers to break the mass-redshift degeneracy~\cite{Messenger:2011gi,Messenger:2013fya,DelPozzo:2015bna,Wang:2020xwn,Chatterjee:2021xrm,Jin:2022qnj,Dhani:2022ulg}. While these approaches allow cosmological inference without EM counterparts, the resulting constraints are generally limited by poor redshift information and large sky localization uncertainties. Most dark sirens cover sky areas of order $10^{3}\,\mathrm{deg}^{2}$, which amplifies the redshift uncertainty, and degeneracies with orbital inclination further degrade the inferred luminosity distance. As a result, although dark sirens are more numerous than bright sirens, they typically provide weaker constraints on cosmological parameters compared to the single confirmed bright siren with identified EM counterparts. At present, owing to the scarcity of bright sirens and the relatively poor localization of dark sirens, standard siren measurements with the LIGO--Virgo--KAGRA network are not yet precise enough to resolve the Hubble tension. Therefore, future progress will likely rely on either increasing the number of bright siren events or improving the localization accuracy of dark sirens. In this work, we focus on strategies to increase the number of detectable bright siren candidates and the accuracy of distance measurement.

In the coming decades, space-based gravitational-wave observatories such as LISA~\cite{LISA:2017pwj}, Taiji~\cite{Hu:2017mde,Ruan:2018tsw}, and TianQin~\cite{TianQin:2015yph,TianQin:2020hid} will open the millihertz frequency window, enabling observations of a rich population of massive black hole binaries (MBHBs) and extreme mass-ratio inspirals out to redshifts $z \sim 20$~\cite{Klein:2015hvg,LISACosmologyWorkingGroup:2022jok,Babak:2017tow}. Among these sources, MBHB mergers are particularly promising candidates for bright sirens, as they are expected to produce EM counterparts in radio, optical, and infrared bands~\cite{Tamanini:2016zlh}. Such counterparts may be detectable by next-generation electromagnetic facilities, including the Square Kilometer Array (SKA)~\cite{SKA}, the Extremely Large Telescope (ELT)~\cite{EELT}, and the Rubin Observatory~\cite{Rubin}. As a result, LISA is anticipated to deliver a substantial population of bright sirens, offering a powerful avenue for precision measurements of the Hubble constant and tests of modified gravity at cosmological distances~\cite{Tamanini:2016zlh,Wang:2019ryf,Wang:2021srv,Mangiagli:2023ize,LISACosmologyWorkingGroup:2019mwx,Zhan:2025jqg}. Realizing this potential, however, critically depends on the accuracy of GW parameter estimation, in particular the sky localization and luminosity distance measurements, which are essential for identifying the electromagnetic counterparts and host galaxies of MBHB mergers. 

A growing body of literature has demonstrated that orbital eccentricity can significantly enhance GW parameter estimation, particularly in constraining the source distance and sky position, in studies of ground-based detectors such as LIGO and in the decihertz frequency band~\cite{Yang:2022tig,Yang:2022iwn,Yang:2022fgp,Yang:2023zxk,Yang:2024vfy,Sun:2015bva,Ma:2017bux,Pan:2019anf}.
This is mainly because eccentricity introduces higher harmonics into the gravitational wave signal, providing additional information that breaks degeneracies between angular parameters and the luminosity distance. These results naturally motivate an investigation of eccentricity effects in the millihertz band relevant to LISA. Meanwhile, previous studies have shown that MBHBs may retain non-negligible orbital eccentricities when they enter the LISA frequency band. They have demonstrated that environmental dynamical processes, such as those operating in stellar, gaseous, and triple interaction channels, can substantially influence the eccentricity evolution of these binaries~\cite{Roedig:2011rn,Bonetti:2018tpf,Zrake:2020zkw,Valli:2024nbj,Munoz:2018tnj,Siwek:2023rlk}. In dense stellar environments, stars interacting with MBHBs can carry away its orbital energy and angular momentum, leading to orbital hardening and an increase in eccentricity. In gaseous environments, eccentricity is initially driven upward by asymmetric torques exerted by the circumbinary disc, while subsequent gas accumulation around the black holes leads to a saturation of this growth~\cite{Roedig:2011rn,Bonetti:2018tpf,Zrake:2020zkw,Valli:2024nbj,Munoz:2018tnj,Siwek:2023rlk}. And in the triple interaction channel, complex three-body encounters can further pump the eccentricity to even higher values, potentially reaching $\gtrsim 0.9$ upon entering the LISA band~\cite{Bonetti:2018tpf}. These results highlight the necessity of properly accounting for eccentricity in the parameter estimation. At the same time, neglecting eccentricity can lead to significantly larger systematic errors in the inferred parameters~\cite{Martel:1999tm,Favata:2013rwa,Favata:2021vhw,Cho:2022cdy,GilChoi:2022waq,Yang:2026mam}. Most previous analyses have examined the impact of eccentricity only on a single event. In~\cite{Mikoczi:2012qy}, for example, the authors found that for a $\sim10^7 M_{\odot}$ MBHB, the angular resolution can be improved by about one order of magnitude when the source has an initial eccentricity of $e_0=0.6$. A comprehensive population-level forecast assessing how eccentricity affects the full ensemble of sources detectable by LISA is needed. The study in~\cite{Yang:2022fgp} investigated this problem in the decihertz band; however, analogous investigations in the millihertz regime are still lacking. Filling this gap is important because millihertz sources are expected to emit electromagnetic counterparts and thus can be used as bright sirens; achieving high-precision parameter estimation in this band can therefore substantially improve our constraints on the expansion history of the Universe. In addition to their impact on individual resolvable sources, recent studies have also begun to explore the role of orbital eccentricity in shaping the gravitational wave background in the LISA band~\cite{Zhao:2025hqd,Raidal:2024odr,Sah:2025dmv}, which remains an important direction for future work.

In this work, following the methodology developed in Refs.~\cite{Klein:2015hvg,Tamanini:2016zlh,LISACosmologyWorkingGroup:2019mwx}, we construct simulated GW catalogs for LISA observations of massive black hole binaries. 
Specifically, we adopt semianalytic MBHB population models (PopIII, Q3d, and Q3nod) to describe the merger rates and the distributions of intrinsic source parameters, including masses, redshifts, and orbital eccentricities, thereby generating realistic mock catalogs for a 5-year LISA mission. Using these simulated datasets, we investigate how orbital eccentricity improves GW parameter estimation for MBHBs observed by LISA, with a particular focus on sky localization and luminosity distance measurements, which are quantified using the Fisher matrix formalism with eccentric inspiral waveforms. Such improvements are expected to increase the number of detectable bright siren candidates by facilitating the identification of EM counterparts and host galaxies. We further assess how the enlarged bright siren candidates leads to tighter constraints on cosmological parameters, including the Hubble constant, as well as enhanced precision in tests of general relativity through GW propagation effects over cosmological distances.

This paper is organized as follows. In Sec.~\ref{sec:typical}, we describe the eccentric waveform model and the Fisher matrix framework used for parameter estimation. Using representative events, we demonstrate how varying eccentricity improves the measurement of luminosity distance and sky position, highlighting the differences between light-seed and heavy-seed populations. In Sec.~\ref{sec:catalog}, we generate mock LISA catalogs based on semianalytic population models and assess the overall improvement in localization and distance accuracy due to eccentricity. In Sec.~\ref{sec:cosmo}, we translate these improvements into forecasts for cosmological constraints using bright sirens. Finally, we conclude our results in Sec.~\ref{sec:conclusion} followed by some discussions.

\section{Fisher-based parameter estimation of the typical binaries \label{sec:typical}}
\subsection{The eccentric waveform and the antenna response}
We adopt the nonspinning, inspiral-only EccentricFD waveform approximant available in {\sc LALSuite}~\cite{lalsuite} and generate the eccentric waveforms using {\sc PyCBC}~\cite{alex_nitz_2024_10473621}. This waveform approximant is based on the enhanced post-circular (EPC) model~\cite{Huerta:2014eca}.  In the limit of vanishing eccentricity, this model consistently reduces to the TaylorF2 post-Newtonian (PN) waveform at 3.5PN order~\cite{Buonanno:2009zt}.   

The frequency-domain eccentric waveform can be written as~\cite{Huerta:2014eca},
\begin{equation}\label{eccfd}
\tilde{h}(f)=-\sqrt{\frac{5}{384}}\frac{\mathcal{M}_c^{5/6}}{\pi^{2/3}d_L}f^{-7/6}\sum_{\ell=1}^{10}\xi_{\ell}\left(\frac{\ell}{2}\right)^{2/3}e^{-i\Psi_{\ell}} \,.
\end{equation}
The phase associated with the $\ell$th harmonic is given by 
\begin{equation}
\Psi_{\ell}=2\pi f t_c-\ell \phi_c+\left(\frac{\ell}{2}\right)^{8/3} \frac{3}{128 \eta v_{\rm ecc}^5} \sum_{n=0}^{7}a_n v_{\rm ecc}^n \,,
\end{equation}
where $t_c$ and $\phi_c$ denote the coalescence time and phase, respectively, and $a_n$ are the standard 3.5PN phase coefficients.
In the circular limit $e_0=0$, the waveform recovers the TaylorF2 model and only the dominant quadrupole mode ($\ell=2$) contributes. For nonzero eccentricity, the coefficients $\xi_{\ell}$ are functions of $e_0$ and angular parameters $P_{\rm ang}= \{ \iota,\theta,\phi,\psi,\beta \}$~\cite{Yunes:2009yz}. The quantity $v_{\rm ecc}$ denotes the eccentricity-modified velocity parameter. Relative to the circular-orbit velocity $v=(\pi M f)^{1/3}$, it is defined as $v_{\rm ecc}(f;e_0)=g(f;e_0)(\pi M f)^{1/3}$, where the function $g(f;e_0)$ encapsulates the eccentricity corrections and is expanded consistently up to ${\cal O}(e_0^8)$. Its explicit form can be found in Eq.~(13) of Ref.~\cite{Huerta:2014eca}. The waveform includes harmonics up to $\ell=10$, corresponding to a self-consistent expansion in eccentricity to ${\cal O}(e^8)$ in both the amplitude and the phase~\cite{Yunes:2009yz}. In total, the nonspinning, inspiral-only eccentric waveform considered here depends on eleven parameters, $P=\{\mathcal{M}_c, \eta,d_L, \iota,\theta, \phi,\psi,t_c, \phi_c, e_0, \beta \}$, where $\mathcal{M}_c$ is the chirp mass, $\eta$ is the symmetric mass ratio, $d_L$ is the luminosity distance, $\iota$ is the inclination angle, $(\theta,\phi)$ specify the sky location of the source, $\psi$ is the polarization angle, and $(t_c, \phi_c)$ denote the time and phase at merger. In addition to the circular TaylorF2 waveform parameters, $e_0$ denotes the initial orbital eccentricity, defined at a reference frequency $f_0$, while $\beta$ represents the azimuthal component of inclination angles (longitude of ascending nodes axis). Eccentricity induces more harmonics to the waveform beyond the dominant quadrupole mode. These additional harmonics introduce nontrivial couplings between the luminosity distance and the angular parameters, thereby helping to break the degeneracies between these parameters. Our eccentric waveform model is valid only for low-to-moderate eccentricities, up to $e \simeq 0.4$. Accordingly, in constructing the mock observations used in this work, we limit the orbital eccentricities to this maximum value. Recent studies using the same semianalytical approach~\cite{Bonetti:2018tpf} have suggested that a substantial fraction of MBHBs could have higher eccentricities, $e > 0.4$. To mimic realistic scenarios as closely as possible while remaining within the validity range of our waveform, we present in Appendix~\ref{app:C} a test in which eccentricities exceeding $0.4$ are capped at $0.4$, while the remainder follows the distribution in~\cite{Bonetti:2018tpf}. In this appendix, we repeat the parameter estimation, including sky localization, luminosity distance inference, and cosmological constraints, using these capped-eccentricity mock data.

In this work, we consider the space-borne detector LISA. Due to the long inspiral duration of MBHBs in the LISA band, the orbital motion of the detector must be taken into account. Following Ref.~\cite{Rubbo:2003ap}, we model LISA with an arm length $L = 2.5 \times 10^{6}\,{\rm km}$. The corresponding antenna pattern functions are defined as
\begin{align}\label{pfunction}
F_{+}(t) &= \frac{1}{2}\left( \cos(2\psi)\, D_{+}(t) - \sin(2\psi)\, D_{\times}(t) \right), \\
F_{\times}(t) &= \frac{1}{2}\left( \sin(2\psi)\, D_{+}(t) + \cos(2\psi)\, D_{\times}(t) \right).
\end{align}
For the inspiral regime, the explicit expressions of $D_{+,\times}$ are adopted in the low-frequency approximation given in Refs.~\cite{Rubbo:2003ap,Ruan:2020smc}. 

For eccentric binaries, the gravitational wave signal contains multiple harmonics of the orbital frequency $F(t)$. The instantaneous frequency of the $\ell$th harmonic is $f_{\ell}(t)=\ell F(t)$. Therefore, a given Fourier frequency $f$ in different harmonics corresponds to different emission times, and hence to different detector orientations along the LISA orbit. As a result, the antenna pattern functions generally differ for different harmonics. As shown in Eq.~(\ref{eccfd}), the EccentricFD waveform implemented in {\sc LALSuite} is constructed by summing over all harmonics in the frequency domain. To consistently incorporate the time and frequency dependent detector response induced by LISA’s orbital motion, we modify {\sc LALSuite} to extract each harmonic component $\tilde{h}_{\ell}(f)$ separately. This treatment allows us to associate each harmonic with the appropriate antenna pattern functions evaluated at its corresponding emission time. To establish the time–frequency relation $t(f)$ for eccentric binaries, we adopt the dominant quadrupole frequency as a reference and numerically solve the phase evolution of eccentric orbits. The response functions for each harmonic mode $\tilde{h}_{\ell}(f)$ are then evaluated as $F_{+,\times}[t(f \rightarrow 2f/\ell)]$. This frequency-dependent modulation of the detector response provides additional information beyond the quasicircular case and plays an important role in breaking the degeneracy between the luminosity distance and angular parameters.

\subsection{Parameter estimation using the Fisher information matrix}
To estimate the uncertainties of the parameters in the waveform, we employ the Fisher information matrix for GWs~\cite{Cutler:1994ys}:
\begin{equation}
\Gamma_{ij}=\left(\frac{\partial h}{\partial P_i},\frac{\partial h}{\partial P_j}\right)\,,
\label{eq:Gamma}
\end{equation}
where $P_i$ denotes one of the 11 waveform parameters. In the circular case, the initial eccentricity $e_0$ and $\beta$ are absent, leaving 9 parameters. The inner product is defined as
\begin{equation}
(a,b) =4\int_{f_{\rm min}}^{f_{\rm max}}\frac{\tilde {a}^*(f)\tilde{b}(f)+\tilde {b}^*(f)\tilde a(f)}{2S_n(f)}df\,.
\label{eq:innerp}
\end{equation}
The sensitivity $S_n(f)$ is adopted from Ref.~\cite{Babak:2021mhe} for LISA. We set the lower and upper frequency cutoffs to $f_{\rm min}=10^{-4}\,{\rm Hz}$ and $f_{\rm max}=\min\left(0.1,f_{\rm ISCO}\right)\,{\rm Hz}$, respectively. Here $f_{\rm ISCO}$ denotes the GW frequency at the innermost stable circular orbit, given by $f_{\rm ISCO}= c^3 (6\sqrt{6}\pi G M)^{-1}$, where $M$ is the total mass. The signal-to-noise ratio (SNR) of a GW event is defined as
\begin{equation}
\rho^2=(h,h) \,,
\end{equation}
and throughout this work we impose an SNR threshold of $\rho=8$.

The covariance matrix of the parameters is given by the inverse of the Fisher matrix, $C_{ij}=(\Gamma^{-1})_{ij}$. The corresponding $1\sigma$ uncertainty on the parameter $P_i$ is then obtained as $\Delta P_i=\sqrt{C_{ii}}$. In this study, we primarily focus on the uncertainties in the luminosity distance $\Delta d_L$ and the sky localization area $\Delta\Omega$, with the latter being defined as
\begin{equation}
\Delta \Omega =2 \pi |\sin(\theta)|  \sqrt{C_{\theta \theta}C_{\phi \phi}-C_{\theta \phi}^2}.
\end{equation} 
The uncertainty in the luminosity distance directly impacts the precision of cosmological parameter inference, while the sky localization accuracy determines the feasibility of identifying electromagnetic counterparts and, consequently, the number of detectable standard siren events. We compute the partial derivatives of the waveform with respect to the parameters numerically using a central finite-difference scheme, $\partial \tilde h/\partial P_i \simeq [\tilde h(f,P_i+dP_i)-\tilde h (f,P_i-dP_i)]/2dP_i$ with $dP_i = 10^{-n_i}$, where $n_i$ is chosen individually for each parameter. The value of $n_i$ is optimized to ensure numerical convergence of the derivatives and the stability of the Fisher matrix. As a consistency check of our implementation, we compute the Fisher matrix using the EccentricFD waveform in the circular limit $e_0=0$ and verify that the resulting parameter uncertainties are consistent with those obtained using the TaylorF2 waveform. This validation paves the way for the usage of EccentricFD waveform with nonvanishing eccentricity.

\subsection{Mocking up typical binaries of MBHBs}

To clearly illustrate the improvement in parameter estimation when eccentricity is included in gravitational waves detectable by LISA, we construct mock observations of two representative MBHB systems based on the population models of Ref.~\cite{Klein:2015hvg,Tamanini:2016zlh}. Following the methodology in~\cite{Klein:2015hvg,Tamanini:2016zlh}, we employ semianalytical models to track the cosmic evolution of black hole masses, spins, and their surrounding gas. In this work, we consider three population models, which differ in the black hole seeding mechanism and the inclusion (or absence) of merger delays:
\begin{itemize}
\item[\textbullet] \textit{Pop3:} a light-seed model with delays included, in which BHs form from population III (PopIII) stars (the variant without delays is not considered here);
\item[\textbullet] \textit{Q3d:} a heavy-seed model with delays included, in which MBHs form from the collapse of protogalactic disks;
\item[\textbullet] \textit{Q3nod:} a heavy-seed model without delays between galaxy and BH mergers, representing an optimistic scenario with a higher event rate.
\end{itemize}

\begin{figure}[htbp]
    \centering
    \includegraphics[width=0.45\textwidth]{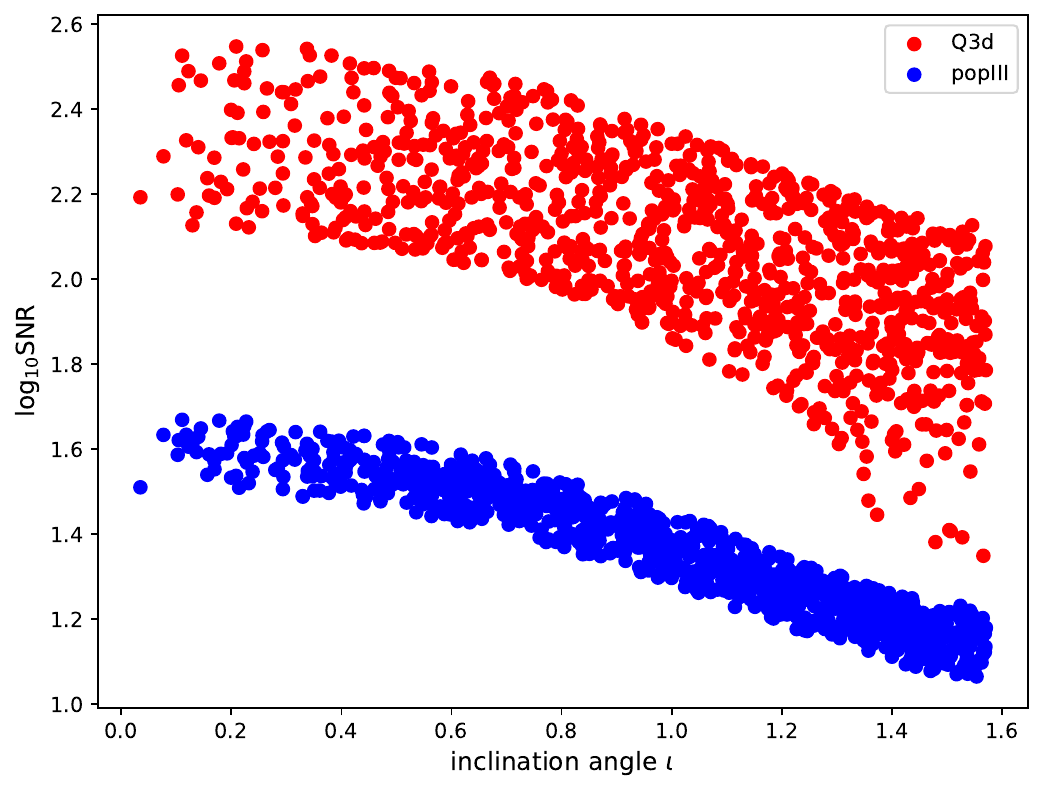}
    \caption{SNR of the two typical binaries with LISA when $e_0=0$. ``PopIII'' correponds to $ (m_1,m_2)=(10^{2.7},10^{2.6})~M_{\odot}$ at $z=1$, and ``Q3d'' is $ (m_1,m_2)=(10^{5.5},10^{5.3})~M_{\odot}$ at $z=4.3$.}
    \label{fig:snr}
\end{figure}

For the illustrative examples in this section, we select typical events from the ``Pop3'' and ``Q3d'' models. The ``Q3nod'' model is omitted here, as we have verified that the presence or absence of merger delays does not qualitatively affect the conclusions regarding the impact of eccentricity. We mock up the typical events for each scenario by choosing component masses corresponding to the peak values of their respective mass distributions. For the PopIII model, we consider the binary with masses $ (m_1,m_2)=(10^{2.7},10^{2.6})~M_{\odot}$ at a redshift of $z=1$. Although the redshift distribution for PopIII sources peaks at higher values ($z \simeq 8$), we select $z=1$ to ensure a sufficient SNR for the Fisher matrix analysis. For the Q3d model, we select the binary with $ (m_1,m_2)=(10^{5.5},10^{5.3})~M_{\odot} $ at the peak of its redshift distribution $z=4.3$. The luminosity distance of each binary is computed assuming a flat $\Lambda$CDM cosmology with $H_0=67.26\,{\rm km\,s^{-1}\,Mpc^{-1}}$ and $\Omega_m=0.315$~\cite{Rosenberg:2022sdy}. The chirp mass $\mathcal{M}_c$ and symmetric mass ratio $\eta$ are then derived from the component masses. All waveform parameters are defined in the detector frame. To account for the impact of source orientation, we generate 1000 independent realizations of the angular parameters $P_{\rm ang}$ for each binary, sampling from isotropic distributions with $\cos\theta\in[-1,1]$, $\phi\in[0,2\pi]$, $\cos\iota\in[0,1]$, $\psi\in[0,2\pi]$, $\beta\in[0,2\pi]$. Without loss of generality, the coalescence time and phase are fixed to $t_c=0$ and $\phi_c=0$. Since the EPC waveform model is validated for initial eccentricities up to $e_0=0.4$~\cite{Huerta:2014eca}, we consider four discrete values of the initial eccentricity at the reference frequency $f_0=10^{-4}\,\rm Hz$, namely, $e_0=0, 0.1, 0.2$, and $0.4$. In total, this procedure yields $2\times4\times1000=8000$ mock realizations. Figure.~\ref{fig:snr} displays the dependence of the SNR on the inclination angle for the two representative binaries in the circular case observed by LISA. As expected, the SNR decreases with increasing inclination angle. For a given binary, we find that the SNR does not vary significantly with eccentricity.

\subsection{Parameter estimation results for typical binaries}
We calculate the Fisher matrix for the two typical binaries described above, covering all 8000 simulated cases. In our analysis, we focus primarily on the estimation errors of the luminosity distance and the source localization. These two parameters are critical for the cosmological application of gravitational waves: precise sky localization allows for identifying electromagnetic counterparts, while accurate distance inference directly reduces the uncertainty in the measurement of cosmological parameters, such as the Hubble constant~\cite{Yang:2022fgp}.

\begin{figure*}[htbp]
    \centering
    \includegraphics[width=0.8\textwidth]{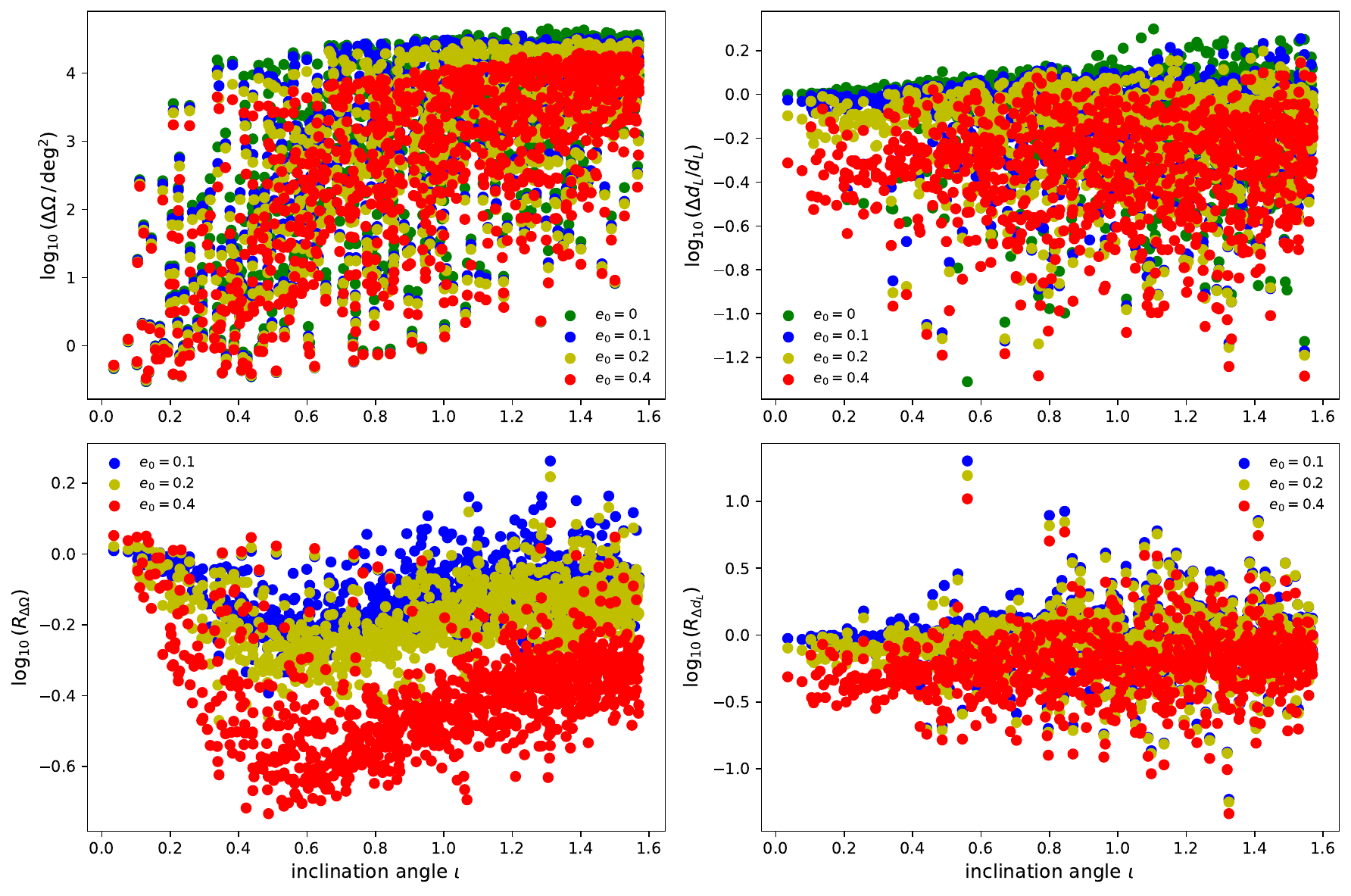}
    \caption{Error and improvement ratio $R$ for luminosity distance and sky localization for the typical Q3d MBHB observed by LISA.}
    \label{fig:heavyseed}
\end{figure*}

\begin{figure*}[htbp]
    \centering
    \includegraphics[width=0.8\textwidth]{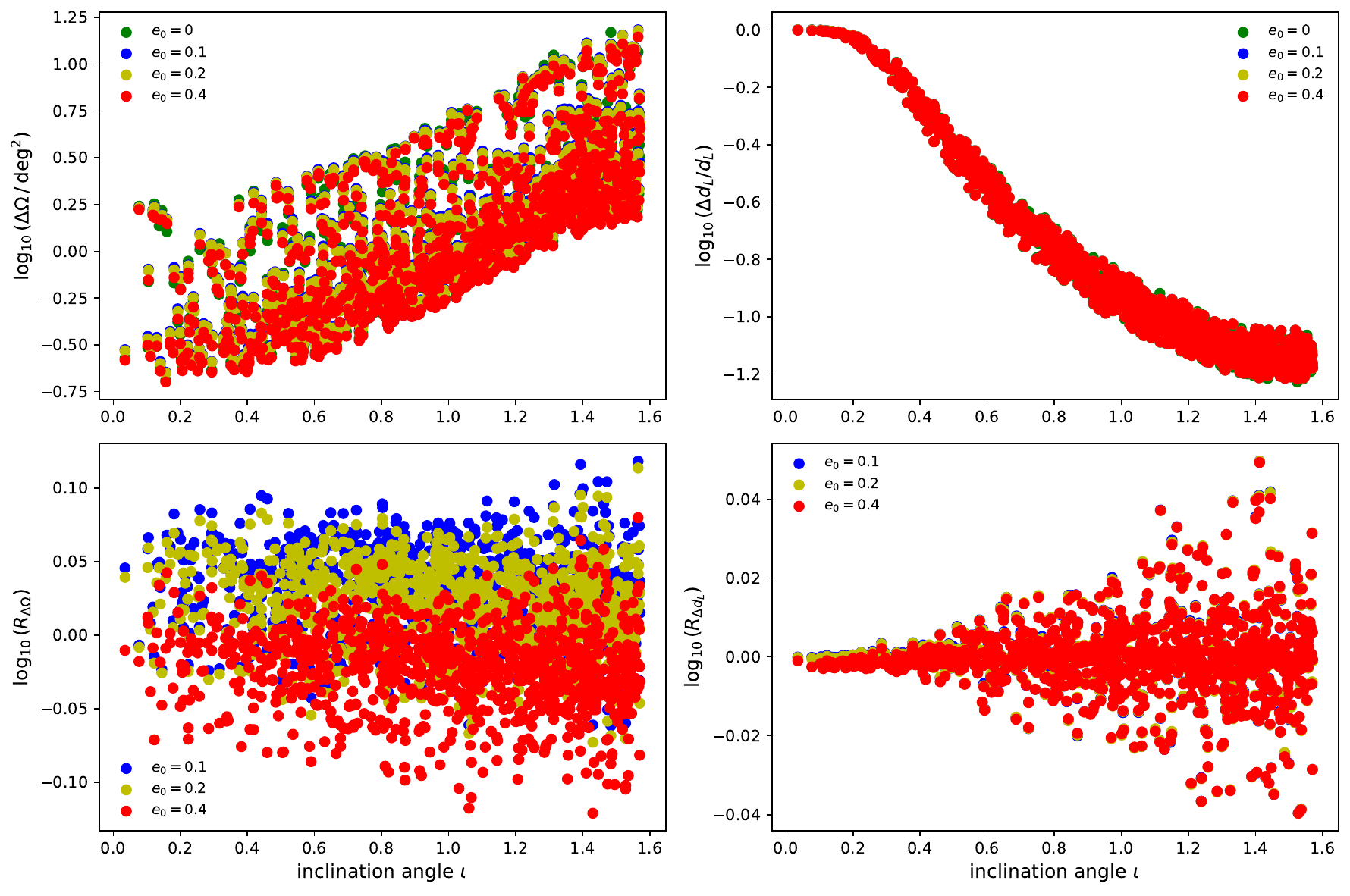}
    \caption{Error and improvement ratio $R$ for luminosity distance and sky localization for the typical PopIII MBHB observed by LISA.}
    \label{fig:lightseed}
\end{figure*}

Figures.~\ref{fig:heavyseed} and \ref{fig:lightseed} display the estimation errors for luminosity distance and sky localization, along with the improvement factor induced by eccentricity, for the Q3d (heavy seed) and PopIII (light seed) models, respectively. Since the SNR and parameter degeneracies are strongly dependent on the source orientation, we present our results as a function of the inclination angle $\iota$. To quantify the impact of eccentricity, we define the improvement ratio:
\begin{equation}
R_{\Delta P}=\frac{\Delta P|_{e_0= \rm nonzero}}{\Delta P|_{e_0=0}}\,,
\end{equation} 
where $\Delta P$ represents the error of a given parameter P. A ratio $R<1$ indicates an improvement in parameter estimation when eccentricity is included, with smaller values corresponding to more significant improvements.

For the typical event in the Q3d model, as shown in Fig.~\ref{fig:heavyseed}, the inclusion of eccentricity leads to substantial improvements in both sky localization and luminosity distance inference. These improvements become increasingly pronounced as the initial eccentricity $e_0$ grows. In the most favorable source orientations, the sky localization accuracy can be enhanced by up to $\sim 0.7$ orders of magnitude, while the precision of luminosity distance estimation improves by approximately one order of magnitude for $e_0 = 0.4$. The improvements depend sensitively on the inclination angle. For sky localization, we find that sources with an intermediate inclination angle experience the largest relative improvement induced by eccentricity.
This is mainly because, for nearly face-on systems, the luminosity distance and inclination angle are most strongly degenerate, such that the inclusion of eccentricity can in principle improve parameter estimation. However, in this regime, the sky localization accuracy is already relatively high in the circular case, leaving limited room for further improvement from eccentricity. Conversely, for nearly edge-on systems, although the degeneracy among parameters is relatively weak, the sky localization accuracy in the circular case is comparatively poor. As a result, the inclusion of eccentricity can still lead to a noticeable improvement in parameter estimation, albeit smaller than that achieved for binaries with intermediate inclination angles. For luminosity distance inference, when the eccentricity is 0.4, the uncertainty is reduced by an average of 0.5 orders of magnitude, with the best-case improvement reaching up to 1.4 orders of magnitude. We find that, for certain events, the inclusion of eccentricity leads to degraded sky localization and distance estimation. This arises because introducing eccentricity enlarges the parameter space, and the additional waveform information is not always sufficient to break parameter degeneracies, thereby increasing uncertainties for individual sources. Nevertheless, the full sample still exhibits improved parameter constraints.
In contrast, for the typical event in the PopIII model, the impact of eccentricity is found to be negligible, as illustrated in Fig.~\ref{fig:lightseed}. Even when the initial eccentricity reaches $e_0 = 0.4$, the inferred parameter uncertainties show only marginal differences compared to the quasicircular case. This indicates that, for such lower-mass events, eccentricity does not play a significant role in improving parameter estimation for individual representative events. 

In summary, the analysis of these representative binaries demonstrates that, for heavy MBHB systems characterized by the Q3d model, the inclusion of orbital eccentricity can lead to a noticeable enhancement in parameter estimation accuracy, which is particularly relevant for cosmological applications of bright sirens. The improvement is most significant for events with larger total masses and lower redshifts, corresponding to higher signal-to-noise ratios. For the PopIII model, Fig.~\ref{fig:lightseed} shows that the improvement for the typical mass scale is not prominent. Nevertheless, it should be emphasized that the representative binaries discussed here are chosen to correspond to the peaks of their respective population distributions. Within the full PopIII population, there exists a non-negligible subset of systems with relatively higher masses and lower redshifts, which can achieve sufficiently high SNR. For these events, the inclusion of eccentricity may still yield noticeable improvements in parameter estimation. We will investigate these statistical variations and assess their cumulative impact on the population-level results in the following sections.

\section{Construction of the bright siren candidates catalogs \label{sec:catalog}}
For typical massive black hole binaries, we have seen that orbital eccentricity can significantly enhance sky localization and reduce the uncertainty in distance measurements. In this section, we aim to quantify the improvement in parameter estimation for mock LISA observations, constructed using the same semianalytical population models as in Refs.~\cite{Klein:2015hvg,Tamanini:2016zlh} based on~\cite{Barausse:2012fy,Sesana:2014bea,Antonini:2015cqa,Antonini:2015sza}. In constructing the bright siren candidates catalogs, we adopt a procedure similar to that adopted in Ref.~\cite{Klein:2015hvg,Tamanini:2016zlh} for waveform and parameter estimation. In this work, the inspiral waveform is modeled using the \texttt{EccentricFD} approximant, which includes the effect of orbital eccentricity but is restricted to the inspiral phase. To account for the contribution from merger and ringdown, we apply a correction by using results obtained with inspiral-merger-ringdown (IMR) “PhenomD” waveforms and characterize its impact through a SNR gain factor $R(\rho)=\rho_{\rm IMR}/\rho_{\rm I}$. This SNR gain factor is then consistently applied to correct the Fisher matrix results for the merger and ringdown contributions. The sky localization accuracy scales approximately as $R(\Delta\Omega)\propto [R(\rho)]^{-2}$, while the relative uncertainty in the luminosity distance scales as $R(\Delta d_L/d_L)\propto [R(\rho)]^{-1}$. Importantly, the same correction procedure is applied to both eccentric and quasicircular binaries, ensuring that the relative comparison between the two cases remains unbiased. As a result, this treatment does not affect our conclusions regarding the impact of eccentricity, but instead provides a more realistic estimate of the parameter estimation performance. Moreover, this approach employs a more complete effective waveform description when constructing the bright siren candidate catalogs, thereby enabling a more reliable assessment of the resulting cosmological parameter constraints. 

\subsection{Population of MBHBs and parameter estimation}
To forecast the improvement in multimessenger detections of MBHBs with LISA enabled by orbital eccentricity, we construct merger catalogs using the population models introduced in Sec.~\ref{sec:typical}. From the three models (PopIII,Q3d and Q3nod), we obtain the merger rate and the intrinsic binary properties (masses, orientations, luminosity distance), as well as the host galaxy properties (gas and stellar mass, disk mass). These catalogs form the basis for our parameter estimation and cosmological inference analysis. For each model, we generate 20 catalogs, each representing 5 years of observations. Following the method used for typical events, we randomly assign angular parameters $P_{\rm ang}$ to each event, drawn from uniform and isotropic distributions: $\cos\theta \in [-1,1]$, $\phi \in [0,2\pi]$, $\cos\iota \in [0,1]$, $\psi \in [0,2\pi]$, and $\beta\in[0,2\pi]$. For the orbital eccentricity, we select four discrete initial values at $f_0 = 10^{-4}\,\rm Hz$, namely, $e_0 = 0, 0.1, 0.2, 0.4$. 
Given the simulated MBHB population, we compute the SNR and perform parameter estimation for each GW event using the Fisher matrix formalism to infer the sky localization and luminosity distance uncertainties.

\subsection{Observation of electromagnetic counterparts}
For bright sirens, identifying the host galaxy of each event is necessary to obtain redshift information, which requires observations of EM counterparts. We select all events with $\mathrm{SNR} > 8$ and $\Delta \Omega < 10~\rm deg^2$, corresponding to the typical field-of-view limits of EM surveys.
We consider two observing strategies: the first employs the Rubin Observatory~\cite{Rubin} to detect optical emission from AGNs, and the second utilizes the SKA~\cite{SKA} and ELT~\cite{EELT} to detect radio jets and flares.
For optical detection, quasar-like luminosity (a fraction of the Eddington luminosity) is expected to be triggered during mergers in gas-rich environments. Using the galaxy evolution model described above, we estimate the available gas reservoir and accretion rate. From these, we infer the expected bolometric and apparent luminosities, which determine whether the flares are detectable by Rubin. We adopt $m_{\rm Rubin} = 26$ as the detection threshold~\cite{Tamanini:2016zlh}. For events satisfying
\begin{equation}\label{eq:Rubin_lim}
\begin{split}
m &= 82.5 + BC
- \frac{5}{2}\log_{10}\left(\frac{L_{\mathrm{bol}}}{3.02}\frac{\mathrm{sec}}{\mathrm{erg}}\right) \\
&\quad + 5\log_{10}\left(\frac{d_L}{\mathrm{pc}}\right)
\leq m_{\rm Rubin},
\end{split}
\end{equation}
the sources can be followed up spectroscopically to obtain precise redshift measurements. Here $BC$ is the bolometric correction, and we adopt a fiducial value $BC = 1$. Details of the calculation are provided in Appendix~\ref{app:A}.
For radio detection, the total radio luminosity is given by $L_{\rm radio} = L_{\rm flare} + L_{\rm jet}$. The first term corresponds to dual jets launched near merger, with luminosity
\begin{equation}
L_{\rm flare} = \epsilon_{\rm edd}\epsilon_{\rm radio}\left(\frac{v}{v_{\rm max}}\right)^2 q^2 L_{\rm edd}.
\end{equation}
Here $\epsilon_{\rm edd} = L_{\rm bol}/L_{\rm edd}$ is the Eddington ratio, computed from the gas accretion rate at merger (see Appendix~\ref{app:A}), with a floor imposed at $\epsilon_{\rm edd} = 0.02$~\cite{Moesta:2011bn}. The parameter $\epsilon_{\rm radio}$ is the fraction of EM radiation emitted in the radio band and is set to the fiducial value 0.1~\cite{OShaughnessy:2011nwl}. The factor $(v/v_{\rm max})^2$ accounts for luminosity evolution during inspiral, where $v_{\rm max} = c/\sqrt{3}$ is the circular speed at the innermost stable circular orbit (ISCO) for nonspinning BH binaries, and $q = M_2/M_1 \leq 1$ is the mass ratio.
The second component is the jet launched by accretion through the Blandford–Znajek mechanism~\cite{Blandford:1977ds}, with luminosity
\begin{widetext}
\begin{equation}\label{eq:L_jet}
L_{\rm jet} =
\begin{cases}
10^{42.7}{\rm erg\ s^{-1}}
\left( \dfrac{\alpha}{0.01} \right)^{-0.1}
m_{9}^{0.9}
\left( \dfrac{\dot m}{0.1} \right)^{6/5}
\left( 1 + 1.1 a_1 + 0.29 a_1^2 \right),
& \text{if } 10^{-2} \le \epsilon_{\rm edd} \le 0.3, \\[1.2em]
10^{45.1}{\rm erg\ s^{-1}}
\left( \dfrac{\alpha}{0.3} \right)^{-1}
m_{9}
\left( \dfrac{\dot m}{0.1} \right)
g^{2}
\left( 0.55 f^{2} + 1.5 f a_1 + a_1^2 \right),
& \text{otherwise}.
\end{cases}
\end{equation}
\end{widetext}
We assume the Shakura–Sunyaev viscosity parameter $\alpha=0.1$. Here $m_9 = M_1/(10^9 M_\odot)$, $\dot m = \dot M/(22 m_9 M_\odot\ {\rm yr}^{-1})$, and $a_1$ is the spin of the primary BH. Parameters $f$ and $g$ regulate the angular velocity and azimuthal magnetic field and are set to $f = 1$ and $g = 2.3$ following~\cite{Meier:2000wk}. Although we focus on transient counterparts triggered by mergers, the above jet luminosity also applies to steadily accreting massive BHs. We assume that merger-induced variability does not significantly alter the order of magnitude of Eq.~(\ref{eq:L_jet}).
Assuming isotropic radio emission, the SKA detection criterion is~\cite{OShaughnessy:2011nwl}
\begin{equation}
L_{\rm radio} \geq 4\pi d_L^2(z) F_{\rm min}^{\rm SKA},
\end{equation}
where $F_{\rm min}^{\rm SKA} = \nu_{\rm SKA} F_{\nu,{\rm min}}^{\rm SKA}$, with $\nu_{\rm SKA} \simeq 1.4\ {\rm GHz}$ and $F_{\nu,{\rm min}}^{\rm SKA} \simeq 1\ \mu{\rm Jy}$. Thus, events satisfying
\begin{equation}\label{eq:SKA_Lradio}
\left(\frac{L_{\rm radio}}{\rm erg\ s^{-1}}\right)
\left(\frac{d_L}{\rm cm}\right)^{-2}
\geq 4\pi \times 10^{-18}
\left(\frac{F_{\nu,{\rm min}}^{\rm SKA}}{\mu{\rm Jy}}\right)
\left(\frac{\nu_{\rm SKA}}{\rm GHz}\right)
\end{equation}
are considered detectable.
Radio detection alone cannot provide redshifts; therefore, follow-up with optical/IR facilities is required. We consider the MICADO spectrograph on the ELT, covering 1000–2400 nm~\cite{MICADOTeam:2010oam}. Spectroscopy is feasible down to $m_{\rm ELT,sp} = 27.2$ for a 5-hour exposure, while photometric redshifts can be obtained down to $m_{\rm ELT,ph} = 31.3$. For $m_{\rm gal} \leq m_{\rm ELT,sp}$, spectroscopic redshifts are obtained, while for $m_{\rm ELT,sp} < m_{\rm gal} \leq m_{\rm ELT,ph}$, photometric redshifts are obtained. The host galaxy magnitude is
\begin{equation}\label{eq:ELT_lim}
m_{\rm gal} = 82.5
-\frac{5}{2}\log_{10}\left(\frac{L_k}{3.02}\frac{\rm s}{\rm erg}\right)
+5\log_{10}\left(\frac{d_L}{\rm pc}\right),
\end{equation}
where $L_k$ is computed from the stellar mass assuming $M/L_k = 0.03$.
Any EM counterpart satisfying Eqs.~(\ref{eq:Rubin_lim}),~(\ref{eq:SKA_Lradio}), and~(\ref{eq:ELT_lim}) will be detectable, and the corresponding redshift can be measured.

\subsection{The improvement of localization and distance inference}

In our analysis, we first construct 20 independent mock GW catalogs, each corresponding to 5 years of LISA observations, for each MBHB population model. For every catalog, we consider four discrete initial eccentricities, $e_0 = 0, 0.1, 0.2, 0.4$, and assume that all binaries share the same initial eccentricity (see Appendix~\ref{app:C} for the result of using a realistic eccentricity distribution). For each event, we compute SNR and perform parameter estimation using the Fisher matrix, from which we infer the sky localization uncertainty $\Delta \Omega$ and the relative luminosity distance uncertainty $\Delta d_L / d_L$. We adopt an SNR threshold of 8. Events satisfying both $\mathrm{SNR} > 8$ and $\Delta \Omega < 10~\rm deg^2$ are considered potentially observable by electromagnetic facilities. For these events, we further apply the method described above to assess the detectability of their electromagnetic counterparts. Events that meet all of these criteria are classified as bright siren candidates. Among the 20 realizations for each population model, we select the catalog with the median number of bright siren candidates as a representative case. In this section, we present results for the Q3nod model.

\begin{figure*}[htbp]
    \centering
    \includegraphics[width=0.8\textwidth]{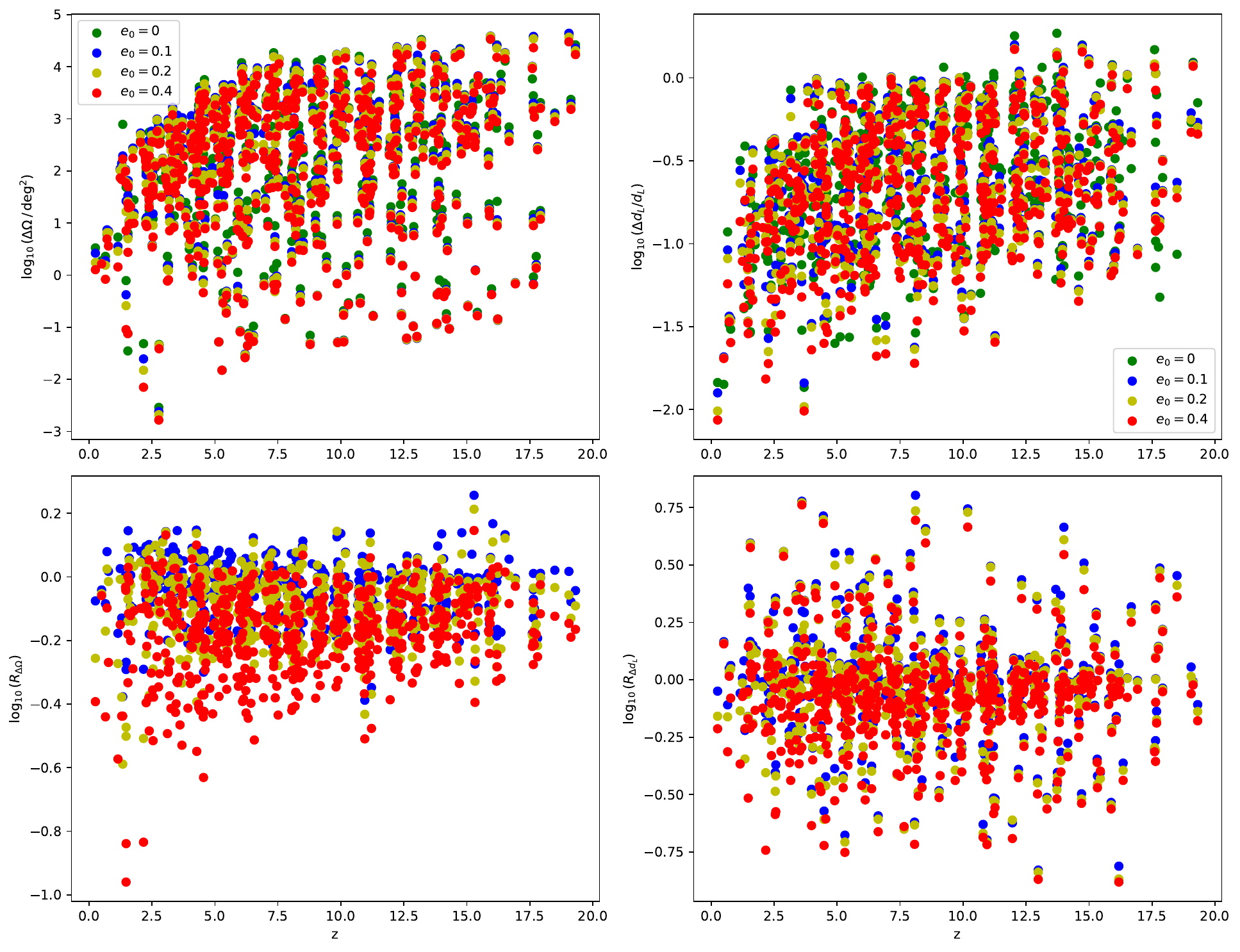}
    \caption{Sky localization uncertainty $\Delta \Omega$ and relative luminosity distance uncertainty $\Delta d_L/d_L$ for events from the Q3nod model observed by LISA over 5 years, for different initial eccentricities. The top panels show the absolute values of $\Delta \Omega$ and $\Delta d_L/d_L$, while the bottom panels display the corresponding improvement factors $R$ relative to the circular case.}
    \label{fig:Q3nod_location}
\end{figure*}

Following the same presentation adopted for the typical binary examples discussed earlier, the top panels display the sky localization uncertainty $\Delta \Omega$ and the relative luminosity distance uncertainty $\Delta d_L/d_L$ as functions of redshift for different initial eccentricities. The bottom panels show the corresponding improvement factors $R$, defined as the ratio of each parameter relative to the circular case. As shown in Fig.~\ref{fig:Q3nod_location}, for the Q3nod model, both sky localization and distance inference are significantly improved when orbital eccentricity is included. The improvement becomes more pronounced for larger initial eccentricities and is particularly strong at lower redshifts. In the most favorable cases, the sky localization accuracy improves by up to $\sim 1$ order of magnitude, while the relative distance uncertainty is reduced by up to $\sim 0.75$ orders of magnitude. These results indicate that sky localization is more sensitive to orbital eccentricity than luminosity distance inference. This behavior is consistent with our findings for typical binaries discussed earlier, where eccentricity provides more information to break degeneracies between angular parameters, leading to a larger relative gain in sky localization accuracy. 

The results for the PopIII and Q3d models are presented in Appendix~\ref{app:B} (see Figs.~\ref{fig:pop3location} and~\ref{fig:Q3dlocation}). The Q3d model exhibits results similar to those of the Q3nod model. In contrast, the overall improvement in the PopIII model is smaller, mainly due to the lower masses and consequently lower SNR of PopIII binaries. Nevertheless, for low-redshift sources ($z \lesssim 10$), where electromagnetic counterpart identification is feasible, eccentricity still leads to a noticeable improvement.

Finally, from these mock MBHB observations, we conclude that orbital eccentricity significantly improves the sky localization of GW sources, thereby increasing the number of events that qualify as bright siren candidates. At the same time, eccentricity reduces the uncertainty in luminosity distance inference, enhancing the cosmological information carried by each individual event. These two effects act synergistically to yield tighter constraints on cosmological parameters, as will be demonstrated in the next section.

\section{Cosmological implications \label{sec:cosmo}}
In this section, we explore the cosmological implications of bright sirens with orbital eccentricity by constructing the GW Hubble diagram and performing parameter inference. Our goal is to quantify the improvement in cosmological parameter constraints enabled by the enhanced sky localization and distance measurements due to orbital eccentricity.

The luminosity distance of a source at redshift $z$, $d_L(z)$, depends on the underlying cosmology and can be written as
\begin{equation}
d_L(z) = c (1+z) \int_0^z \frac{dz'}{H(z')}\,,
\end{equation}
where $H(z)$ is the Hubble parameter describing the expansion rate of the Universe at redshift $z$. In this work, we consider two cosmological models: $\Lambda$CDM and $w$CDM. 

For the $\Lambda$CDM model, the standard cosmological model with a dark energy equation of state $w=-1$, the Hubble parameter is
\begin{equation}
H(z) = H_0 \sqrt{\Omega_m (1+z)^3 + 1 - \Omega_m}\,,
\end{equation}
where $\Omega_m$ is the matter density parameter and $H_0$ is the Hubble constant.

In the $w$CDM model, describing the simplest dynamical dark energy scenario with a constant $w$, the Hubble parameter takes the form
\begin{equation}
H(z) = H_0 \sqrt{\Omega_m (1+z)^3 + (1-\Omega_m) (1+z)^{3(1+w)}}\,.
\end{equation}
Since the $w$CDM model introduces an additional free parameter, bright sirens from LISA alone are insufficient to yield meaningful constraints. To break parameter degeneracies and improve the robustness of the inference, we therefore include complementary CMB data.

In addition to standard cosmology, we also investigate constraints on modified GW propagation using bright sirens. Following Ref.~\cite{LISACosmologyWorkingGroup:2019mwx}, we consider deviations from general relativity arising from a modified friction term in the GW propagation equation,
\begin{equation}
h_A'' + 2 \mathcal{H}[1 - \delta(\eta)] h_A' + k^2 h_A = 0 \,,
\end{equation}
where $\delta(\eta)$ encodes deviations from GR. Such modifications lead to a difference between the GW luminosity distance $d_L^{\rm GW}$ and the electromagnetic luminosity distance $d_L^{\rm EM}$,
\begin{equation}
d_L^{\rm GW}(z) = d_L^{\rm EM}(z) 
\exp\left[ -\int_0^z \frac{dz'}{1+z'} \, \delta(z') \right] \,.
\end{equation}

Following Ref.~\cite{Belgacem:2018lbp}, we adopt a two-parameter phenomenological parametrization,
\begin{equation}\label{eq:para_MG}
\Xi(z) \equiv \frac{d_L^{\rm GW}(z)}{d_L^{\rm EM}(z)} 
= \Xi_0 + \frac{1 - \Xi_0}{(1+z)^n} \,,
\end{equation}
which captures the predictions of a broad class of modified gravity models. In general relativity, one has $\Xi_0 = 1$. Joint GW and EM observations therefore allow direct constraints on $\delta(z)$ or, equivalently, on the parameters $(\Xi_0, n)$.

Throughout this work, we use CMB data from the \texttt{CamSpec} likelihood based on the latest \texttt{NPIPE} PR4 data release from the \emph{Planck} collaboration~\cite{Rosenberg:2022sdy}. Mock bright siren candidate catalogs are generated assuming a flat $\Lambda$CDM fiducial cosmology with $H_0 = 67.26~\mathrm{km\,s^{-1}\,Mpc^{-1}}$ and $\Omega_m = 0.315$~\cite{Rosenberg:2022sdy}. We further fix the present CMB temperature to $T_{\rm CMB} = 2.7255~\mathrm{K}$, the sum of neutrino masses to $\sum m_\nu = 0.06~\mathrm{eV}$, and the effective number of relativistic degrees of freedom to $N_{\rm eff} = 3.044$, consistent with the \emph{Planck} baseline analysis.

\subsection{Hubble diagram of bright Siren candidates}

To model the measurement uncertainty, we include contributions from the instrumental error, weak-lensing error, peculiar velocity of the host galaxy, and redshift measurement error, which can be combined as
\begin{equation}
\sigma_{d_L} = \sqrt{\sigma_{\rm inst}^2 + \sigma_{\rm lens}^2 + \sigma_{\rm pv}^2 + \sigma_{\rm red}^2} \,.
\end{equation}

The first term, $\sigma_{\rm inst}$, represents the instrumental error derived from the Fisher matrix analysis as described above. The second term, $\sigma_{\rm lens}$, accounts for the systematic uncertainty due to weak lensing, a major source of error for high redshift standard sirens. We adopt the fitting formula from \cite{Hirata:2010ba,Tamanini:2016zlh} to evaluate this error,
\begin{equation}
\sigma_{\rm lens}(z) = d_L(z) \times 0.066 \left[ \frac{1-(1+z)^{-0.25}}{0.25} \right]^{1.8} \,,
\end{equation}
and include a delensing factor of two~\cite{Tamanini:2016zlh}.

The third term, $\sigma_{\rm pv}$, captures the uncertainty due to the peculiar velocity of the host galaxy, expressed as \cite{Kocsis:2005vv}
\begin{equation}
\sigma_{\rm pv}(z) = d_L(z) \left[ 1 + \frac{c(1+z)}{H(z) d_L(z)} \right] \frac{\sqrt{\langle v^2 \rangle}}{c} \,,
\end{equation}
where we assume an rms peculiar velocity $ \sqrt{\langle v^2 \rangle}= 500~\rm km/s $.

Finally, the term $\sigma_{\rm red}$ represents the uncertainty associated with redshift measurements. It can be neglected when the redshift is determined spectroscopically. However, for events with photometrically measured redshifts, we estimate the uncertainty as
\begin{equation}
\sigma_{\rm red}(z) = \frac{\partial d_L}{\partial z} \Delta z \,,
\end{equation}
with $\Delta z \simeq 0.03(1+z)$~\cite{Dahlen:2013fea,Ilbert:2013bf}. 

\begin{figure*}[htbp]
    \centering

    \includegraphics[width=0.3\textwidth]{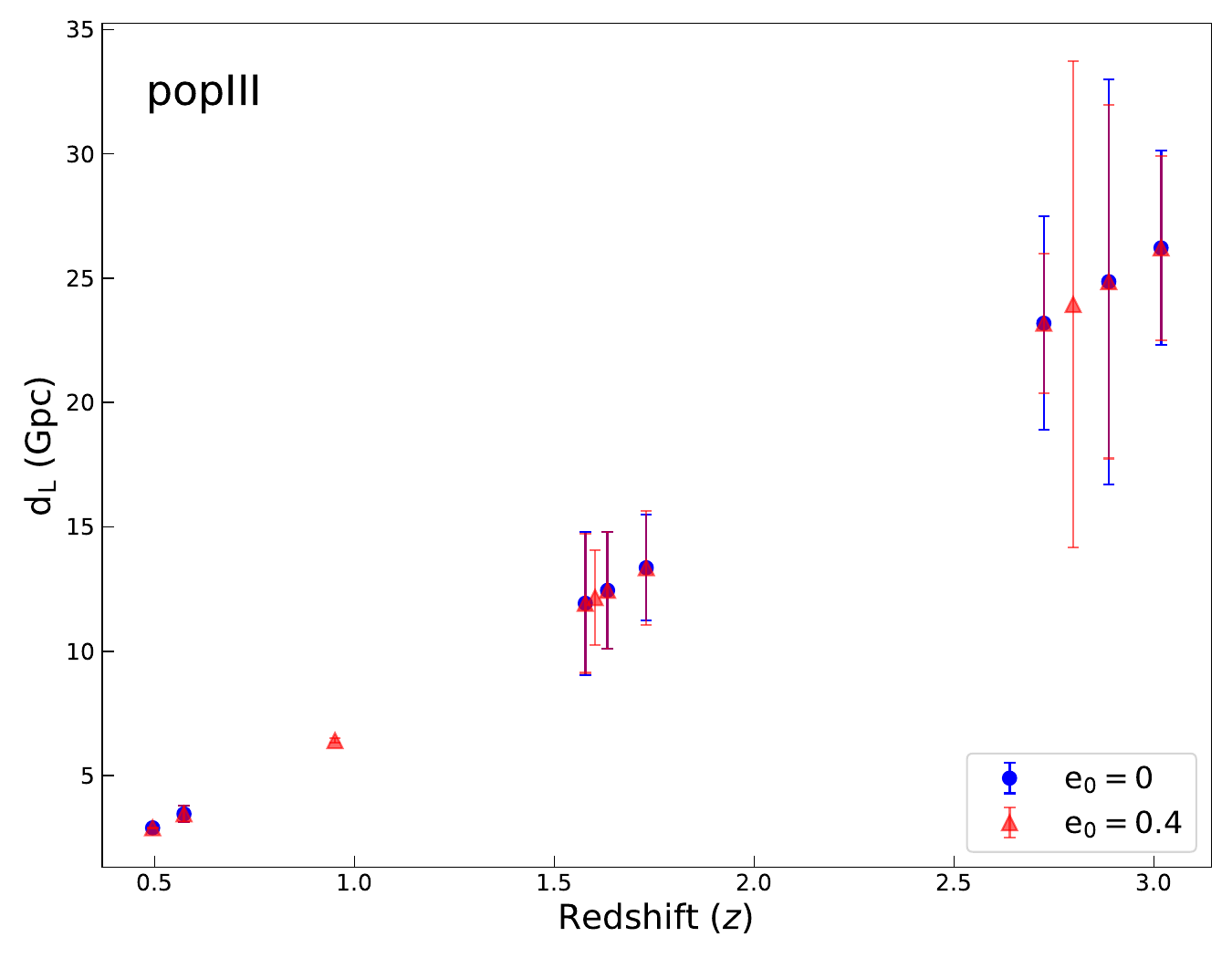}\hfill
    \includegraphics[width=0.3\textwidth]{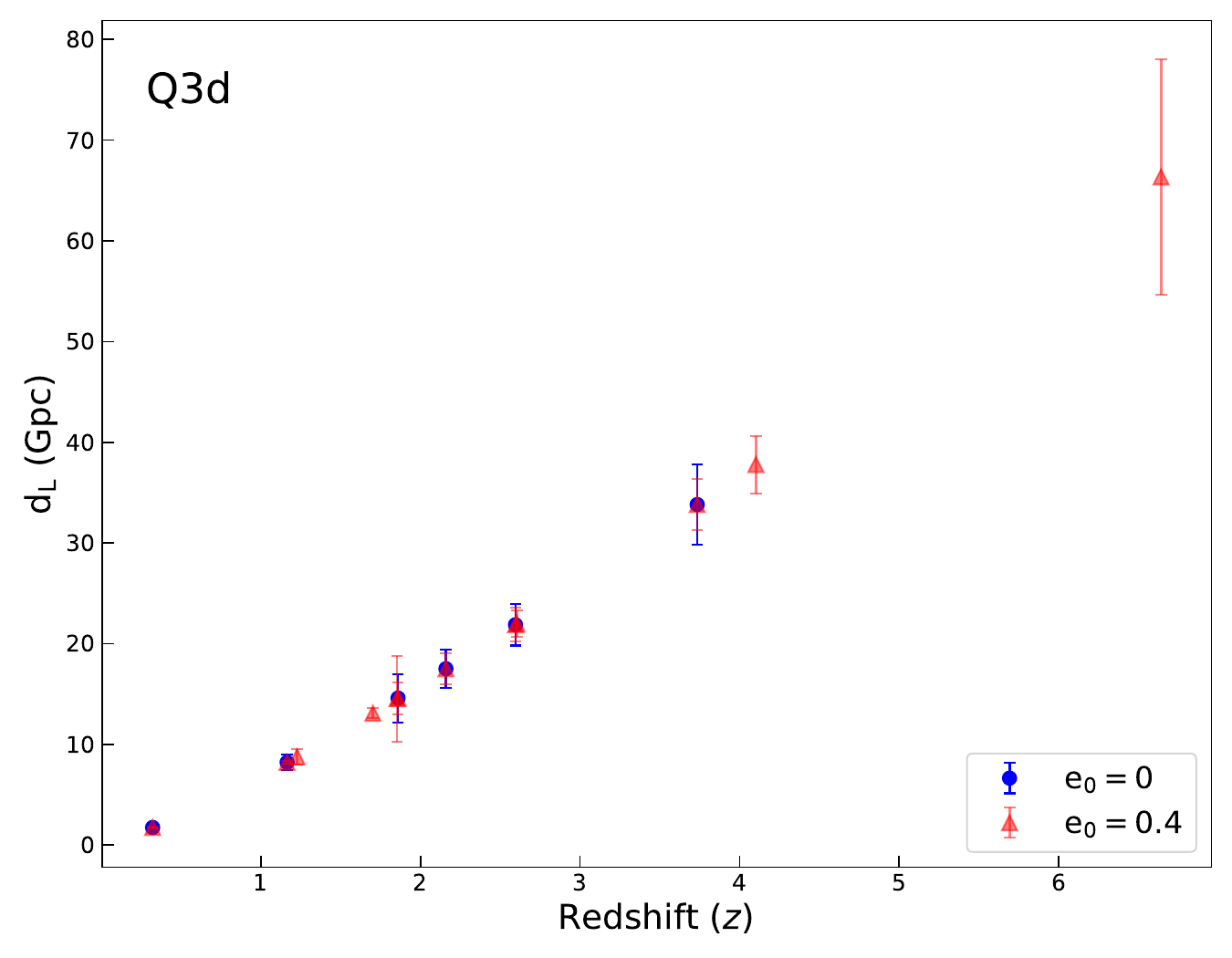}\hfill
    \includegraphics[width=0.3\textwidth]{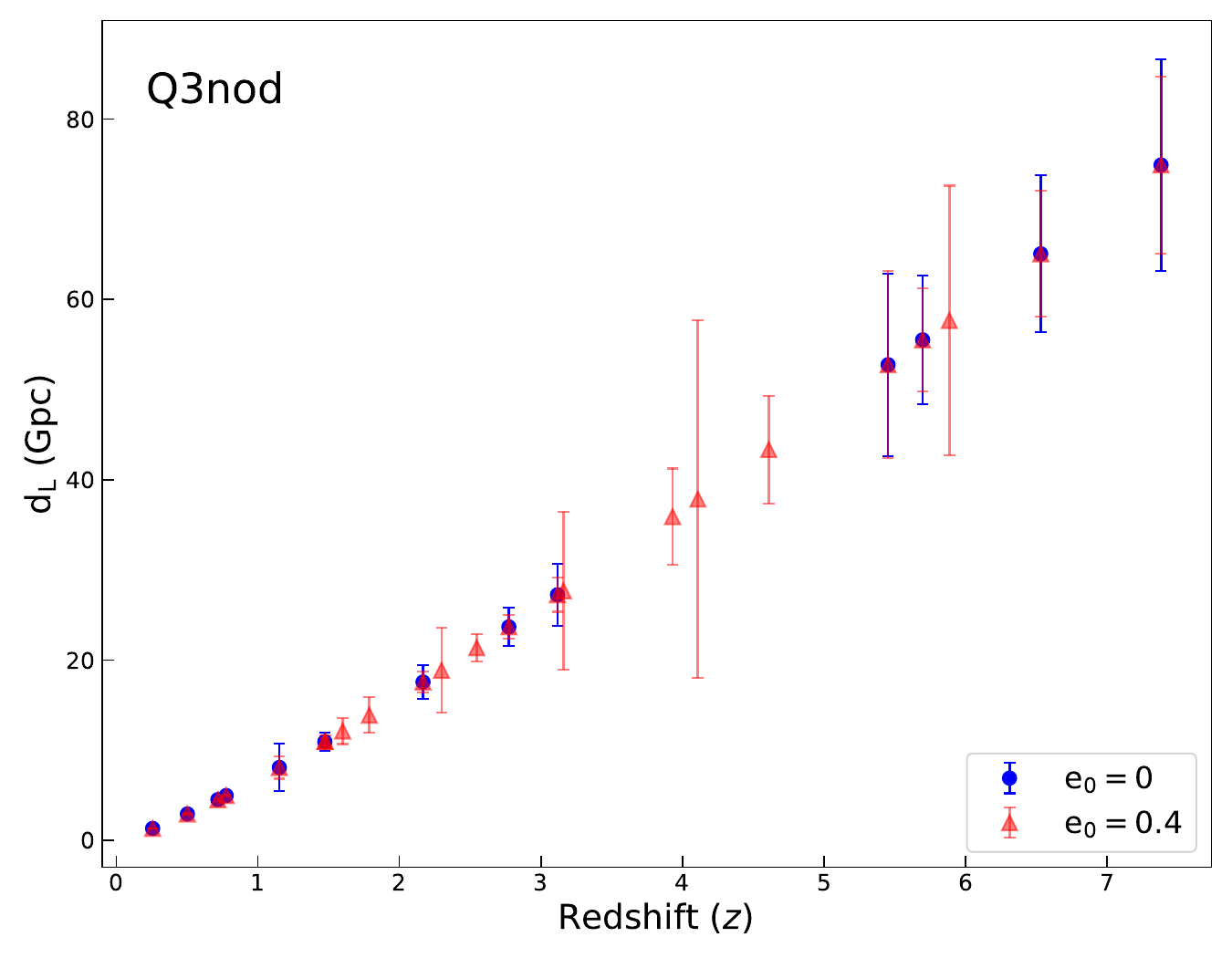}

    \vspace{0.4cm} 

    \includegraphics[width=0.3\textwidth]{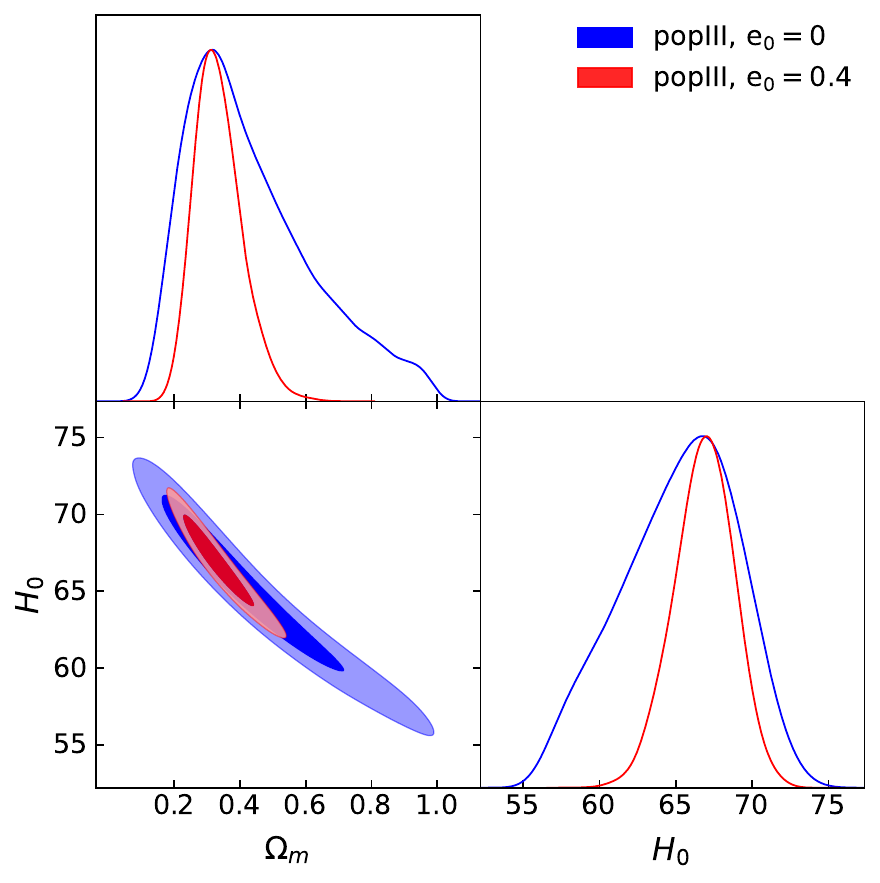}\hfill
    \includegraphics[width=0.3\textwidth]{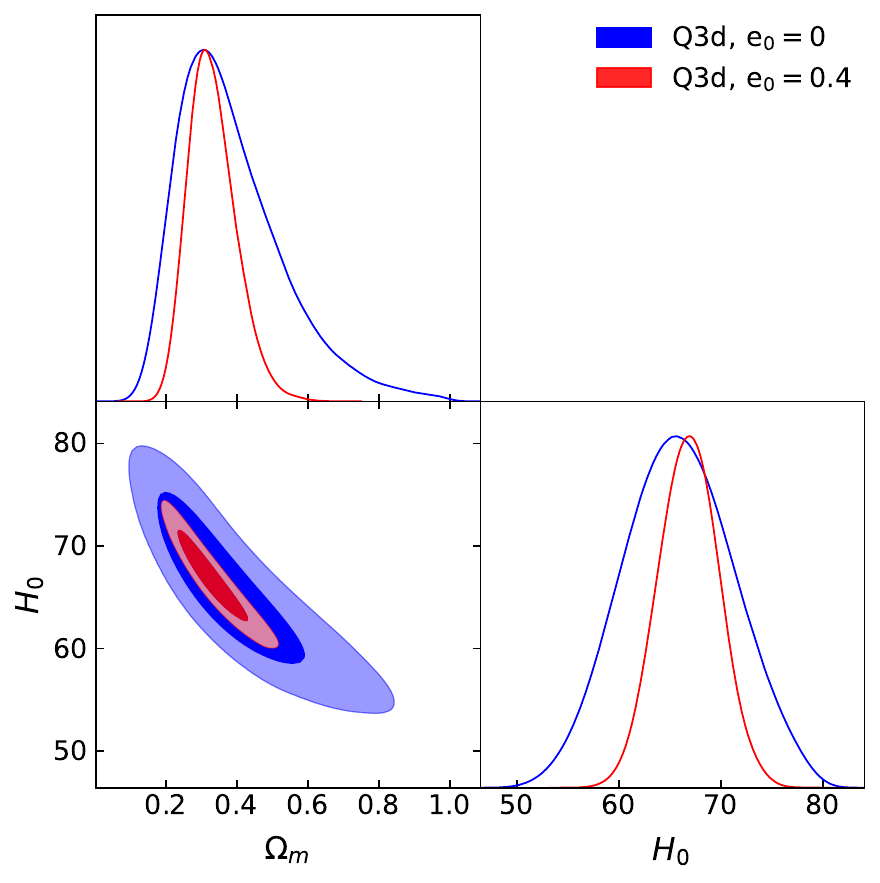}\hfill
    \includegraphics[width=0.3\textwidth]{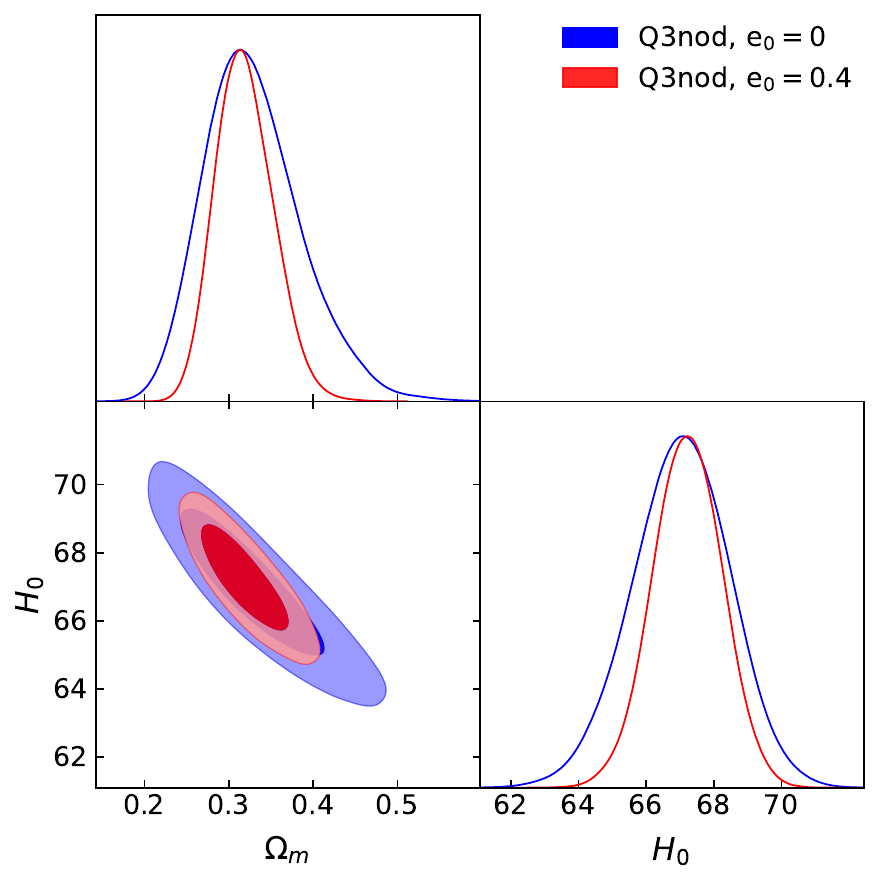}

    \caption{Comparison of the Hubble diagrams (top row) and cosmological constraints (bottom row). Hubble diagram of bright siren candidates observed by LISA over a 5-year mission for each population model with $e_0=0$ and $e_0=0.4$. 
    Constraints of Hubble constant $H_0$ and matter density parameter $\Omega_m$ with $\Lambda$CDM from bright siren candidates observed by LISA in 5 years for each population model with $e_0=0$ and $e_0=0.4$.  The contours correspond to the 68\% and 95\% credible regions.}
    \label{fig:sts cosmo constraints}
    
\end{figure*}

Using these components, we construct the Hubble diagram of bright siren candidates for each population model in Fig.~\ref{fig:sts cosmo constraints}. For each event, the simulated luminosity distance is drawn from a Gaussian distribution centered at the fiducial value $d_L(z)$ with standard deviation $\sigma_{d_L}$.

The upper panels of Fig.~\ref{fig:sts cosmo constraints} show the Hubble diagrams of bright siren candidates for each population model, assuming initial eccentricities of $e_0 = 0$ and $e_0 = 0.4$ for a 5-year LISA observation. We select the catalog with the median number of bright siren candidates among 20 realizations as representative. 

The Hubble diagrams indicate that including orbital eccentricity increases the number of bright siren candidates while simultaneously reducing the luminosity distance uncertainty, both of which are essential for improving cosmological constraints. Since the improvements increase with eccentricity, we present results for $e_0 = 0$ and $e_0 = 0.4$ as representative cases. Specifically, the number of bright siren candidates increases from 8 to 11 for the PopIII model, from 6 to 12 for the Q3d model, and from 13 to 24 for the Q3nod model.

\subsection{Constraints on cosmological models from bright sirens}

Within the fiducial $\Lambda$CDM model, we consider two free cosmological parameters, the Hubble constant $H_0$ and the matter density parameter $\Omega_m$. We perform a Markov chain Monte Carlo (MCMC) analysis to infer the posterior distributions of these parameters using the \texttt{emcee} package~\cite{2013PASP..125..306F}, and independently verify the results with the \texttt{Cobaya} framework~\cite{Torrado:2020dgo,CobayaASCL}.

The cosmological parameter constraints inferred from the three representative catalogs discussed in the previous section are shown in the lower panels of Fig.~\ref{fig:sts cosmo constraints}. These contours correspond to the bright siren Hubble diagrams displayed in the upper panels of the same figure.

As shown in Fig.~\ref{fig:sts cosmo constraints}, including orbital eccentricity leads to a significant improvement in the cosmological parameter inference for all population models. The effect is particularly pronounced for the Q3d model. This can be understood from the fact that, among the three population models, Q3d yields the smallest number of bright siren candidates in the circular case, resulting in relatively weak constraints on cosmological parameters. When eccentricity is taken into account, the number of detectable bright siren candidates in the Q3d model increases significantly, leading to a much larger relative improvement in parameter precision compared to the other two models, which already provide comparatively strong constraints in the circular case. Among the three population models, the Q3nod model consistently yields the tightest cosmological constraints, owing to its larger number of bright siren candidates.

We denote a generic cosmological parameter by $\xi$, and use $\sigma(\xi)$ and $\varepsilon(\xi)$ to represent its absolute and relative uncertainties, respectively, with $\varepsilon(\xi) \equiv \sigma(\xi)/\xi$, where $\sigma(\xi)$ corresponds to the 68\% confidence level.
Quantitatively, for the Hubble constant, the relative uncertainty in the PopIII model decreases from $\varepsilon(H_0)=5.82\%$ for $e_0=0$ to $\varepsilon(H_0)=2.98\%$ for $e_0=0.4$. For the Q3d model, $\varepsilon(H_0)$ is reduced from $8.17\%$ to $4.35\%$, while for the Q3nod model it decreases from $2.18\%$ to $1.52\%$ when eccentricity is included.
An even more pronounced improvement is observed for the matter density parameter. The relative uncertainty $\varepsilon(\Omega_m)$ decreases from $63.17\%$ to $23.81\%$ for the PopIII model, from $48.57\%$ to $21.27\%$ for the Q3d model, and from $18.41\%$ to $10.79\%$ for the Q3nod model as the initial eccentricity increases from $e_0=0$ to $e_0=0.4$. All numerical results are summarized in Table~\ref{tab:error}.

\begin{table*}[htbp]
\centering
\renewcommand{\arraystretch}{1.25} 
\caption{Absolute and relative errors (1$\sigma$) of the cosmological parameters in the $\Lambda$CDM model for different population models observed by LISA with $e_0 = 0$ and $e_0 = 0.4$.}
\label{tab:error}

\begin{tabular}{
    c|
    c@{\hspace{10pt}}c@{\hspace{10pt}}c|
    c@{\hspace{10pt}}c@{\hspace{10pt}}c
}
\toprule
& \multicolumn{3}{c|}{LISA ($e_0 = 0$)} 
& \multicolumn{3}{c}{LISA ($e_0 = 0.4$)} \\
\cline{2-7}
Error
& PopIII & Q3d & Q3nod
& PopIII & Q3d & Q3nod \\
\midrule

$\sigma(\Omega_m)$   
& 0.199 & 0.153 & 0.058
& 0.075 & 0.067 & 0.034 \\

$\sigma(H_0)$        
& 3.917 & 5.494 & 1.463
& 2.005 & 2.929 & 1.022 \\

\midrule

$\varepsilon(\Omega_m)$  
& 0.632 & 0.486 & 0.184
& 0.238 & 0.213 & 0.109 \\

$\varepsilon(H_0)$
& 0.0582 & 0.0817 & 0.0218
& 0.0298 & 0.0435 & 0.0152 \\

\bottomrule
\end{tabular}

\end{table*}

Finally, we summarize the implications of including orbital eccentricity for cosmological parameter estimation with bright sirens. First, eccentricity leads to systematically tighter cosmological constraints, with larger eccentricities leading to greater gains in precision. Second, the improvement is most pronounced for massive and low-redshift events, which typically have higher SNRs and thus benefit more from the inclusion of higher harmonics. Third, for population models with relatively small numbers of bright siren candidates, such as the Q3d model, eccentricity-induced increases in event rates play a crucial role in substantially enhancing the constraining power of bright sirens.

\subsection{Constraints on cosmological models from bright sirens combined with CMB}
In the previous subsection, we demonstrated that orbital eccentricity can significantly improve cosmological parameter constraints within the $\Lambda$CDM model when using bright sirens alone. However, for cosmological models with more than two free parameters, bright sirens by themselves are generally insufficient to fully break parameter degeneracies. In this section, we therefore combine bright siren observations with CMB data to investigate the impact of eccentricity on extended cosmological models. We consider two cosmological models: the $\Lambda$CDM model and the $w$CDM model. As before, the $\Lambda$CDM model is characterized by the Hubble constant $H_0$ and the matter density parameter $\Omega_m$. The $w$CDM model extends $\Lambda$CDM by allowing the dark energy equation-of-state parameter $w_0$ to deviate from $-1$, resulting in three free parameters $(H_0, \Omega_m, w_0)$. For the GW data, we adopt the bright siren candidate catalogs from the Q3d population model, as it represents a physically motivated scenario and allows us to clearly illustrate the impact of orbital eccentricity. We perform MCMC analyses using the \texttt{Cobaya} framework, combining $Planck$ CMB data with bright sirens observed by LISA under two assumptions for the initial orbital eccentricity, $e_0 = 0$ and $e_0 = 0.4$.
\begin{figure}[htbp]
    \centering
    \includegraphics[width=0.45\textwidth]{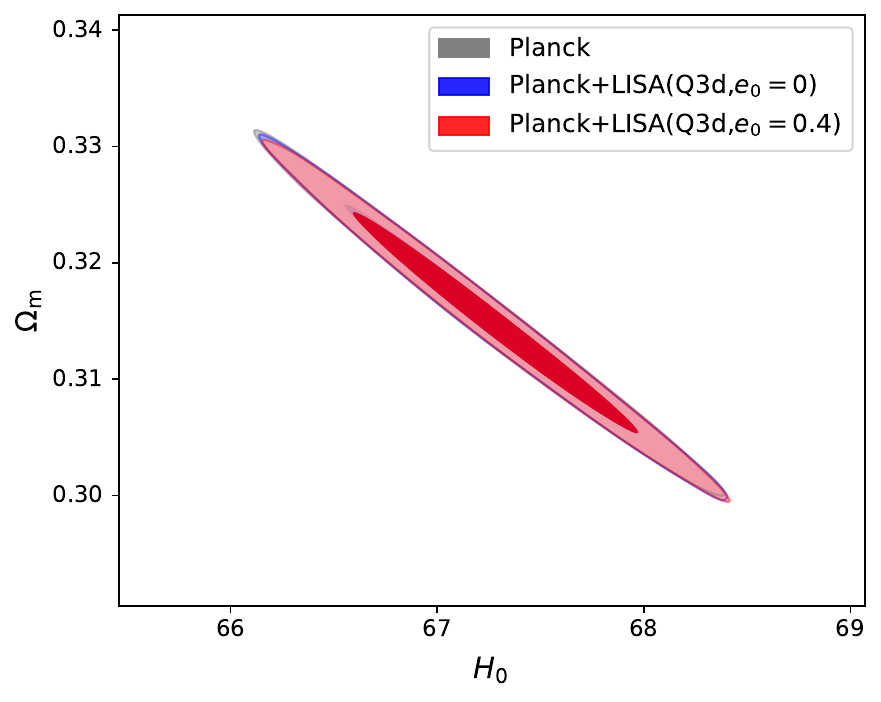}
    \caption{Constraints on $(H_0, \Omega_m)$ in the $\Lambda$CDM model from CMB data alone and from the combination of CMB and LISA bright siren candidates in the Q3d population model, assuming initial eccentricities $e_0=0$ and $e_0=0.4$. The contours correspond to the 68\% and 95\% credible regions.}
    \label{fig:LCDM_Q3d}
\end{figure}

\begin{figure}[htbp]
    \centering
    \includegraphics[width=0.45\textwidth]{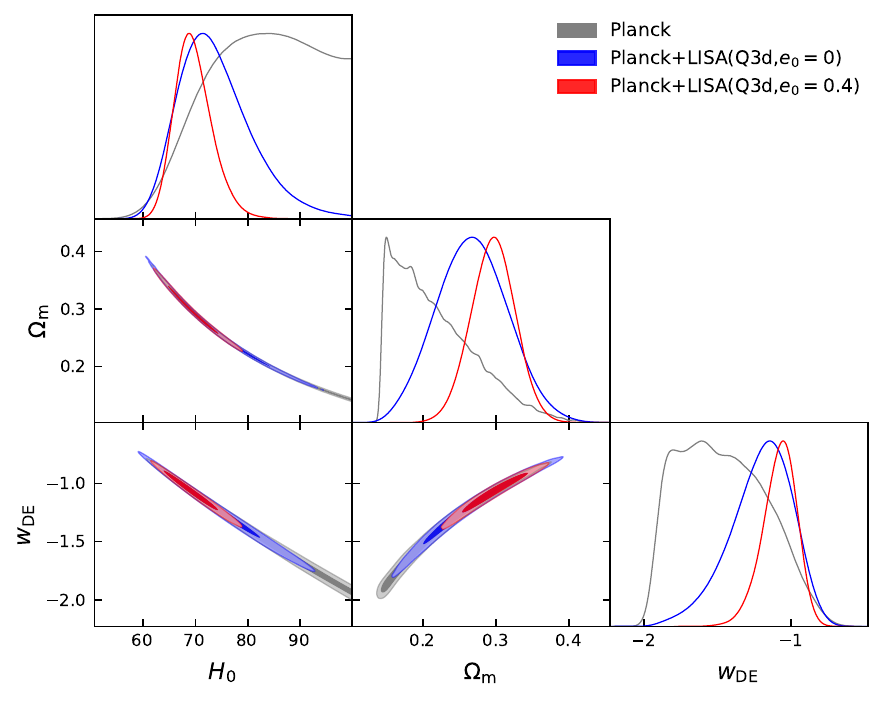}
    \caption{Constraints on $(H_0, \Omega_m, w_0)$ in the $w$CDM model from CMB data alone and from the combination of CMB and LISA bright siren candidates in the Q3d population model, assuming initial eccentricities $e_0=0$ and $e_0=0.4$. The contours correspond to the 68\% and 95\% credible regions.}
    \label{fig:wCDM_Q3d}
\end{figure}

Figure.~\ref{fig:LCDM_Q3d} shows the constraints on $(H_0, \Omega_m)$ in the $\Lambda$CDM model obtained from CMB data alone and from the combination of CMB and LISA bright sirens, for initial eccentricities $e_0=0$ and $e_0=0.4$. We find that CMB observations already provide tight constraints on $\Lambda$CDM parameters, such that the addition of bright sirens leads to only a modest improvement. The further gain induced by including orbital eccentricity is relatively small in this case.

In contrast, the impact of eccentricity becomes considerably more pronounced for the $w$CDM model. Figure.~\ref{fig:wCDM_Q3d} shows the constraints on $(H_0, \Omega_m, w_0)$ in the $w$CDM model from CMB data alone and from the combination of CMB and LISA bright sirens, for initial eccentricities $e_0=0$ and $e_0=0.4$. While CMB data alone cannot tightly constrain all three parameters due to strong degeneracies, the inclusion of bright sirens substantially breaks these degeneracies. Moreover, accounting for orbital eccentricity further enhances the constraining power by increasing both the number of bright siren candidates and the precision of their luminosity-distance measurements. Quantitatively, for the $w$CDM model, the relative uncertainty in the Hubble constant decreases from $\varepsilon(H_0)=14.81\%$ for CMB alone to $\varepsilon(H_0)=10.47\%$ for CMB+LISA with $e_0=0$, and further to $\varepsilon(H_0)=5.4\%$ when eccentricity is included. Similarly, the relative uncertainty in the matter density parameter improves from $\varepsilon(\Omega_m)=17.83\%$ to $15.52\%$ and $9.67\%$, while the uncertainty in the dark energy equation-of-state parameter decreases from $\varepsilon(w_0)=28.87\%$ to $21.45\%$ and $11.77\%$, respectively. We note that our injections assume a $\Lambda$CDM universe ($w=-1$), this choice naturally helps break the degeneracies among $w$CDM parameters, as the data are consistent with a fixed fiducial cosmology.

These results demonstrate that while $\Lambda$CDM parameters are already well constrained by CMB observations, bright sirens, especially when orbital eccentricity is taken into account, play a crucial role in breaking parameter degeneracies and tightening constraints in extended cosmological models such as $w$CDM.

\subsection{Constraints on modified gravity from bright sirens combined with CMB}

In this section, we demonstrate how GW bright sirens combined with CMB data can be used to constrain models of modified gravity (MG). Specifically, we use $Planck$ CMB data to anchor the electromagnetic luminosity distance $d_L^{\rm EM}$, and combine it with LISA bright sirens from the Q3d population model to constrain the modified GW propagation parameter $\Xi_0$ together with the dark energy equation-of-state parameter $w_{\rm DE}$.

To account for the modified GW propagation, we extend the \texttt{Cobaya} framework by incorporating the MG effects into the luminosity distance relation for gravitational waves. We adopt the phenomenological parametrization described in Eq.~(\ref{eq:para_MG}), in which the deviation from general relativity is characterized by the parameter $\Xi_0$ and the exponent $n$. In this work, we fix $n=2.5$, since its impact on the constraints is generally subdominant compared to that of $\Xi_0$, as discussed in~\cite{Belgacem:2018lbp}.

\begin{figure}[htbp]
    \centering
    \includegraphics[width=0.45\textwidth]{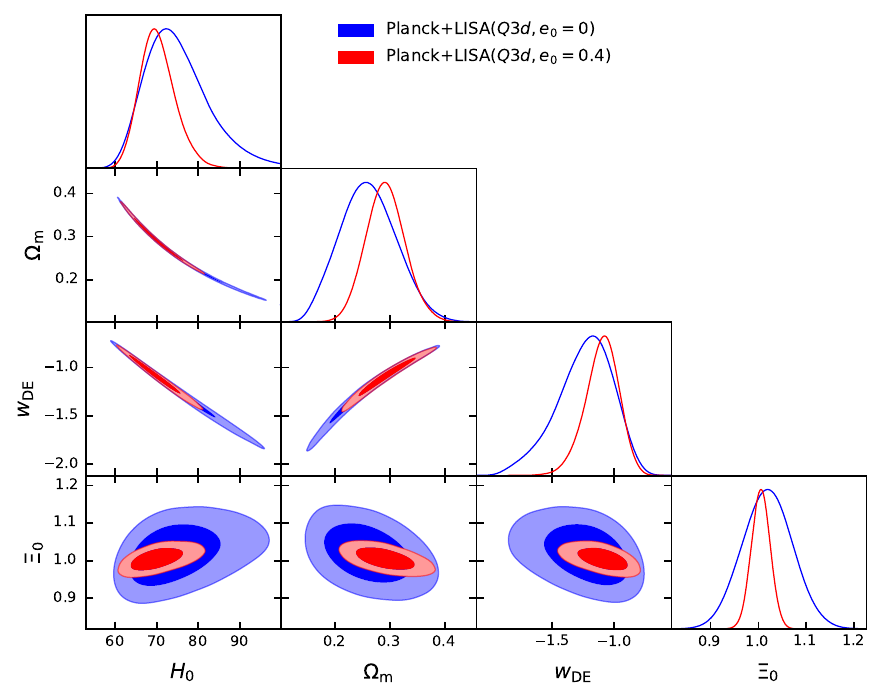}
    \caption{Constraints on $(H_0, \Omega_m, w_{\rm DE}, \Xi_0)$ from the combination of CMB data and LISA bright siren candidates in the Q3d population model, assuming initial eccentricities $e_0=0$ and $e_0=0.4$. The contours correspond to the 68\% and 95\% credible regions.}
    \label{fig:MGQ3d}
\end{figure}

The posterior distributions of the Hubble constant $H_0$, the matter density parameter $\Omega_m$, the dark energy equation-of-state parameter $w_{\rm DE}$, and the modified gravity parameter $\Xi_0$ are shown in Fig.~\ref{fig:MGQ3d} for initial eccentricities $e_0=0$ and $e_0=0.4$.
Figure~\ref{fig:MGQ3d} shows that, for circular binaries ($e_0=0$), the Q3d bright siren candidate catalog constrains the modified gravity parameter to $\Xi_0 = 1.019 \pm 0.0526$. When eccentricity is included ($e_0=0.4$), the constraint is significantly improved to $\Xi_0 = 1.005 \pm 0.0196$. Correspondingly, the relative uncertainty decreases from $\varepsilon(\Xi_0)=5.26\%$ to $\varepsilon(\Xi_0)=1.96\%$. The constraints on the remaining cosmological parameters are also mildly tightened when eccentricity is taken into account.

These results further demonstrate that orbital eccentricity enhances the constraining power of GW bright sirens on modified gravity, primarily by increasing the number of detectable events and improving the precision of luminosity distance measurements.

\section{Conclusions and discussion \label{sec:conclusion}}

In this paper, we have systematically investigated the impact of orbital eccentricity on GW parameter estimation with LISA, and explored its implications for cosmology and fundamental physics. The presence of eccentricity introduces multiple higher harmonics into the GW signal, which provides additional information to break degeneracies among source parameters.

At the level of individual events, we first studied two representative types of massive black hole binaries motivated by semianalytical models, corresponding to heavy seed and light seed formation scenarios. We found that for heavy seed binaries, orbital eccentricity can lead to significant improvements in parameter estimation. Depending on the eccentricity, the inclusion of eccentricity yields optimal gains spanning $\sim 0.4$--$0.7$ orders of magnitude in sky localization and $\sim 0.6$--$1.4$ orders of magnitude in luminosity distance inference. In contrast, for light seed binaries, the improvement is much less pronounced, primarily because their lower masses result in lower SNR, reducing the impact of eccentricity induced higher harmonics on parameter estimation.

In the second part of this work, we constructed mock GW catalogs for three population models (PopIII, Q3d, and Q3nod) observed by LISA over a five-year mission duration. We found that eccentricity systematically improves both sky localization and distance inference for all three models, with the effect being particularly significant for the Q3nod model. We then identified bright siren candidates by selecting events with $\rho>8$ and sky localization accuracy $\Delta \Omega < 10~{\rm deg}^2$, which are likely to be detectable by EM follow-up observations. By further applying EM detectability criteria based on apparent magnitudes, we constructed bright siren candidate catalogs for each population model. For the median realization of each catalog, we found that the number of bright siren candidates increases substantially when eccentricity is included. Specifically, the number of bright siren candidates increases from 8 to 11 for the PopIII model, from 6 to 12 for the Q3d model, and from 13 to 24 for the Q3nod model when the initial eccentricity is set to $e_0=0.4$.

We then investigated the implications of these bright siren candidate catalogs for cosmological parameter estimation. Using bright sirens alone within the $\Lambda$CDM framework, we found significant improvements in the constraints on the Hubble constant $H_0$ and the matter density parameter $\Omega_m$ for all three population models. In particular, the relative uncertainty on $H_0$ is reduced from 5.82\% to 2.98\% and that on $\Omega_m$ from 63.17\% to 23.81\% for the PopIII model, from 8.17\% to 4.35\% and from 48.57\% to 21.27\% for the Q3d model, and from 2.18\% to 1.52\% and from 18.41\% to 10.79\% for the Q3nod model.

Beyond $\Lambda$CDM, we further considered extended cosmological models by combining bright siren observations with $Planck$ CMB data. While CMB data alone already provide tight constraints on $\Lambda$CDM parameters, leaving limited room for improvement, we found that bright sirens play a crucial role in breaking parameter degeneracies in models with additional degrees of freedom, such as the $w$CDM and modified gravity scenarios. In these cases, the inclusion of eccentricity leads to a marked enhancement in constraining power, driven by both the increased number of bright siren candidates and the improved distance precision. In particular, for the $w$CDM model, the degeneracies among $(H_0, \Omega_m, w_0)$ present in CMB constraints are significantly reduced when bright sirens are included, with eccentricity further tightening the constraints. For the Q3d model, the uncertainty in $w_0$ decreases from 28.87\% to 21.45\% and 11.77\% as CMB data are supplemented by LISA and further by eccentric sources. Similarly, for modified gravity models characterized by a modified GW propagation parameter $\Xi_0$, we find that eccentric bright sirens can constrain deviations from general relativity at the percent level, improving the relative uncertainty on $\Xi_0$ from 5.26\% to 1.96\% when going from circular to eccentric binaries.

Overall, our results highlight the importance of orbital eccentricity in LISA science. Eccentricity not only enhances GW parameter estimation at the level of individual events, but also boosts the statistical power of GW catalogs by increasing the number of bright siren candidates. These effects propagate coherently into improved constraints on cosmological and gravitational physics.

Finally, we note several limitations and prospects for future work. The eccentric frequency domain waveform adopted in this study is restricted to initial eccentricities up to $e_0=0.4$. However, previous studies indicate that a non-negligible fraction of LISA sources may exhibit even higher eccentricities, motivating the development of more accurate and broadly applicable eccentric waveform models. In addition, more complex GW features, such as spin precession and higher order modes, may further improve parameter estimation and deserve systematic investigation in future analyses. 

\begin{figure*}[htbp]
    \centering
    \includegraphics[width=0.8\textwidth,height=0.45\textheight]{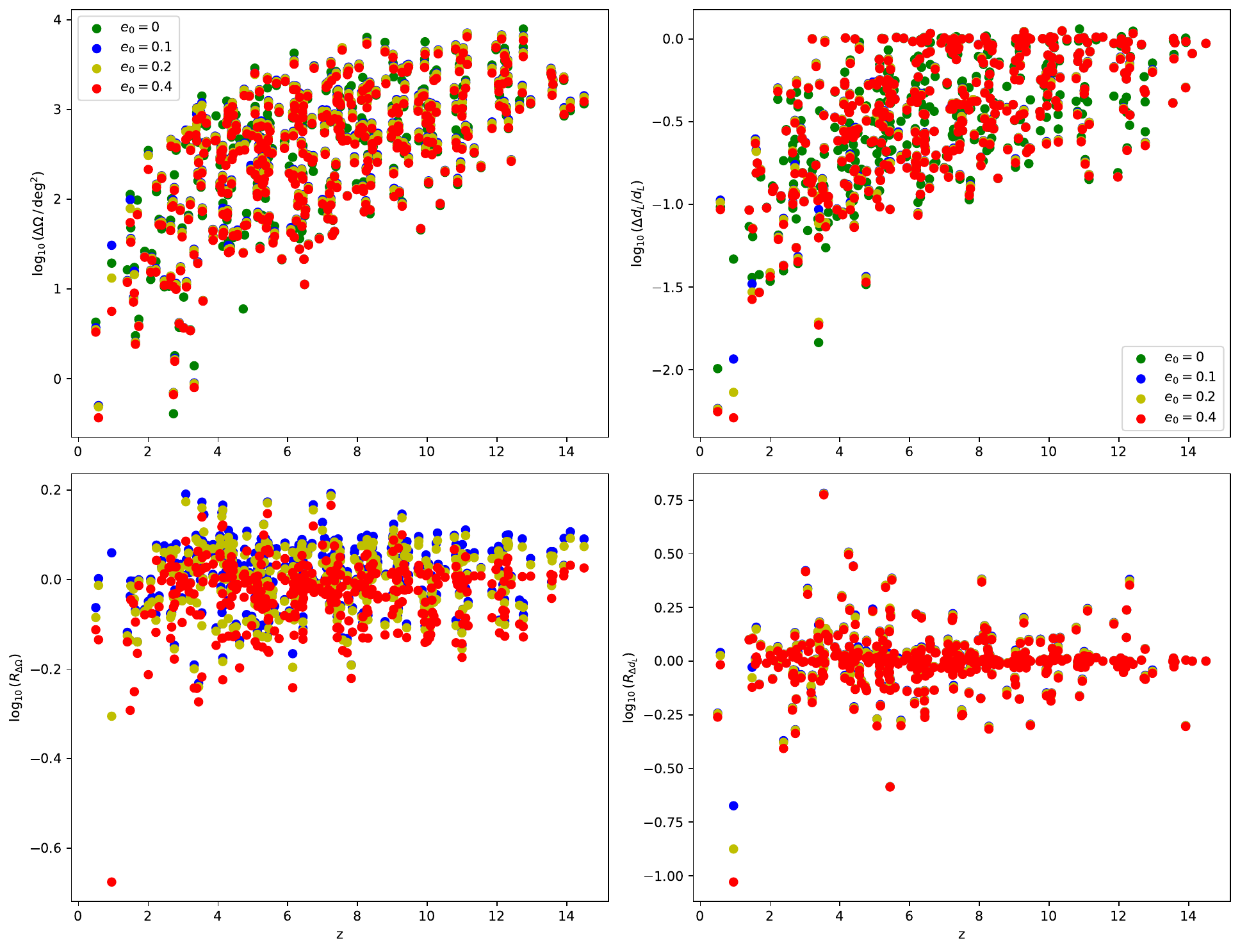}
    \caption{Same as Fig.~\ref{fig:Q3nod_location} in the main text, but for the PopIII model.}
    \label{fig:pop3location}
\end{figure*}

\begin{figure*}[htbp]
    \centering
    \includegraphics[width=0.8\textwidth,height=0.45\textheight]{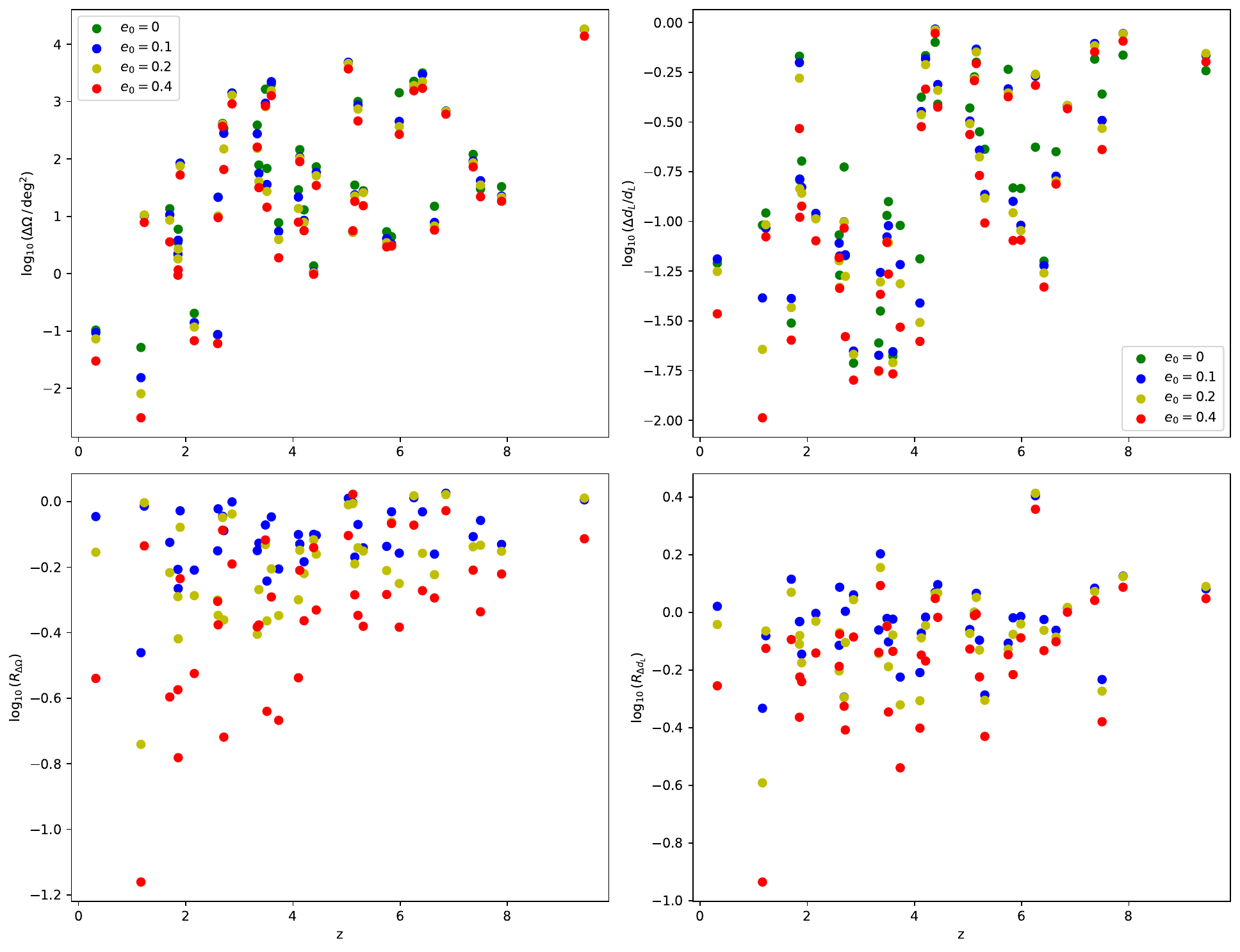}
    \caption{Same as Fig.~\ref{fig:Q3nod_location} in the main text, but for the Q3d model.}
    \label{fig:Q3dlocation}
\end{figure*}

\begin{figure}[htbp]
    \centering
    \includegraphics[width=0.45\textwidth]{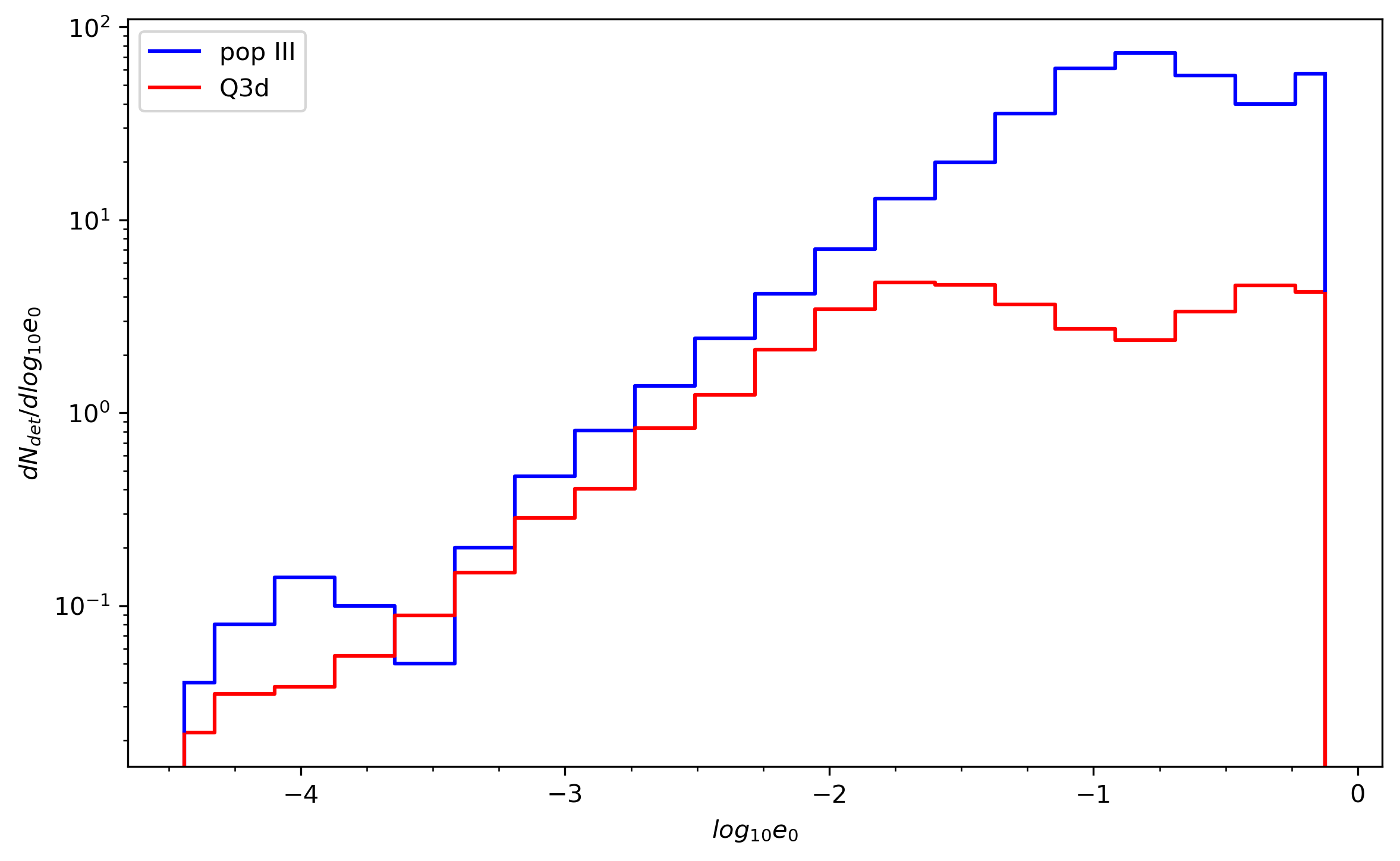}
    \caption{Eccentricity distributions at $f_0 = 10^{-4}\,{\rm Hz}$ for MBHB events with ${\rm SNR} > 8$, constructed from the weighted combination of different evolutionary channels.}
    \label{fig:eccdis}
\end{figure}

\begin{figure*}[htbp]
    \centering
    \includegraphics[width=0.8\textwidth,height=0.43\textheight]{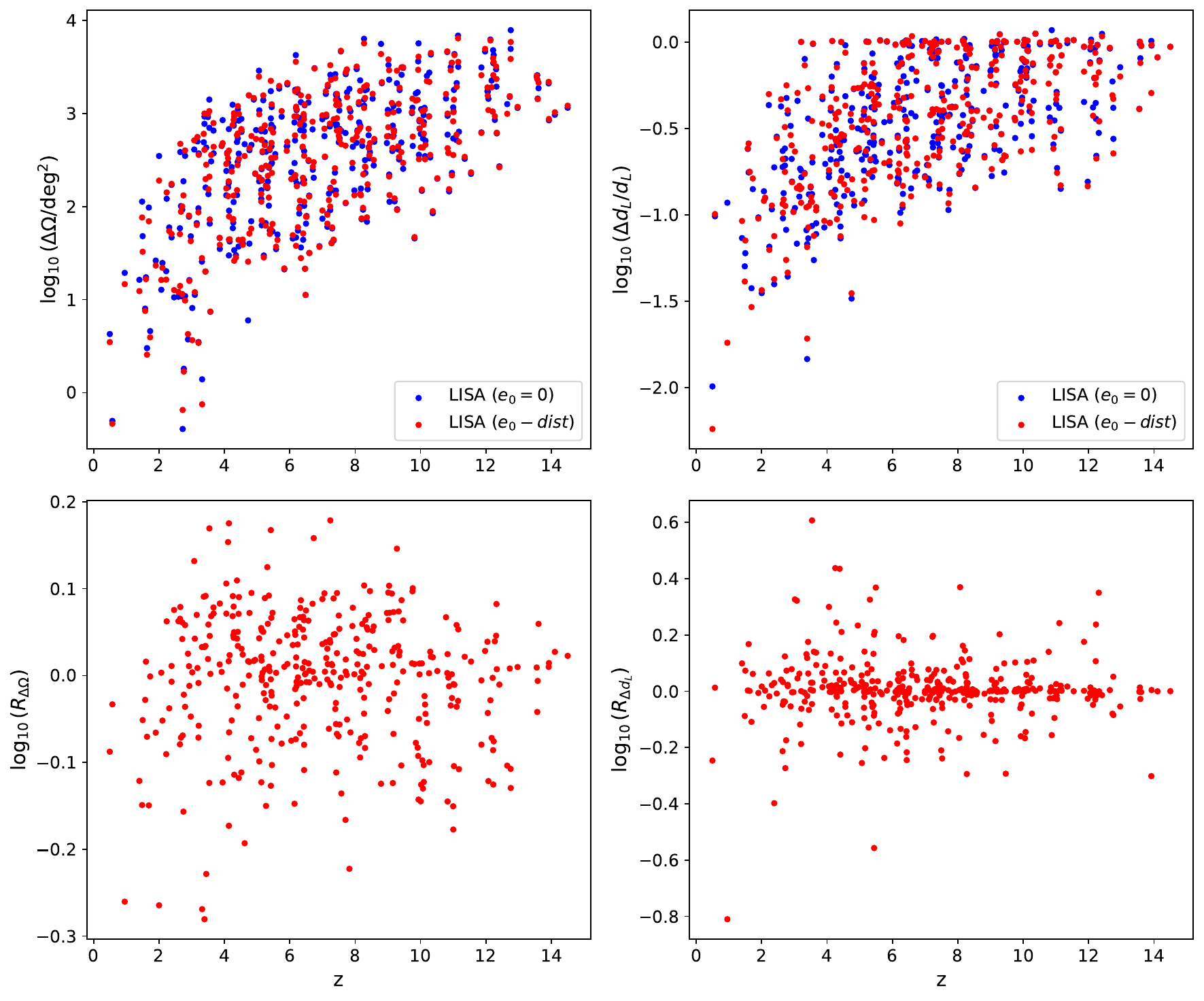}
    \caption{Same as Fig.~\ref{fig:Q3nod_location} in the main text, but for the PopIII model with orbital eccentricities drawn from the distribution shown in Fig.~\ref{fig:eccdis}. The label $e_0$-dist indicates that the eccentricities are sampled from the adopted distribution.}
    \label{fig:pop3elocation}
\end{figure*}

\begin{figure*}[htbp]
    \centering
    \includegraphics[width=0.8\textwidth,height=0.43\textheight]{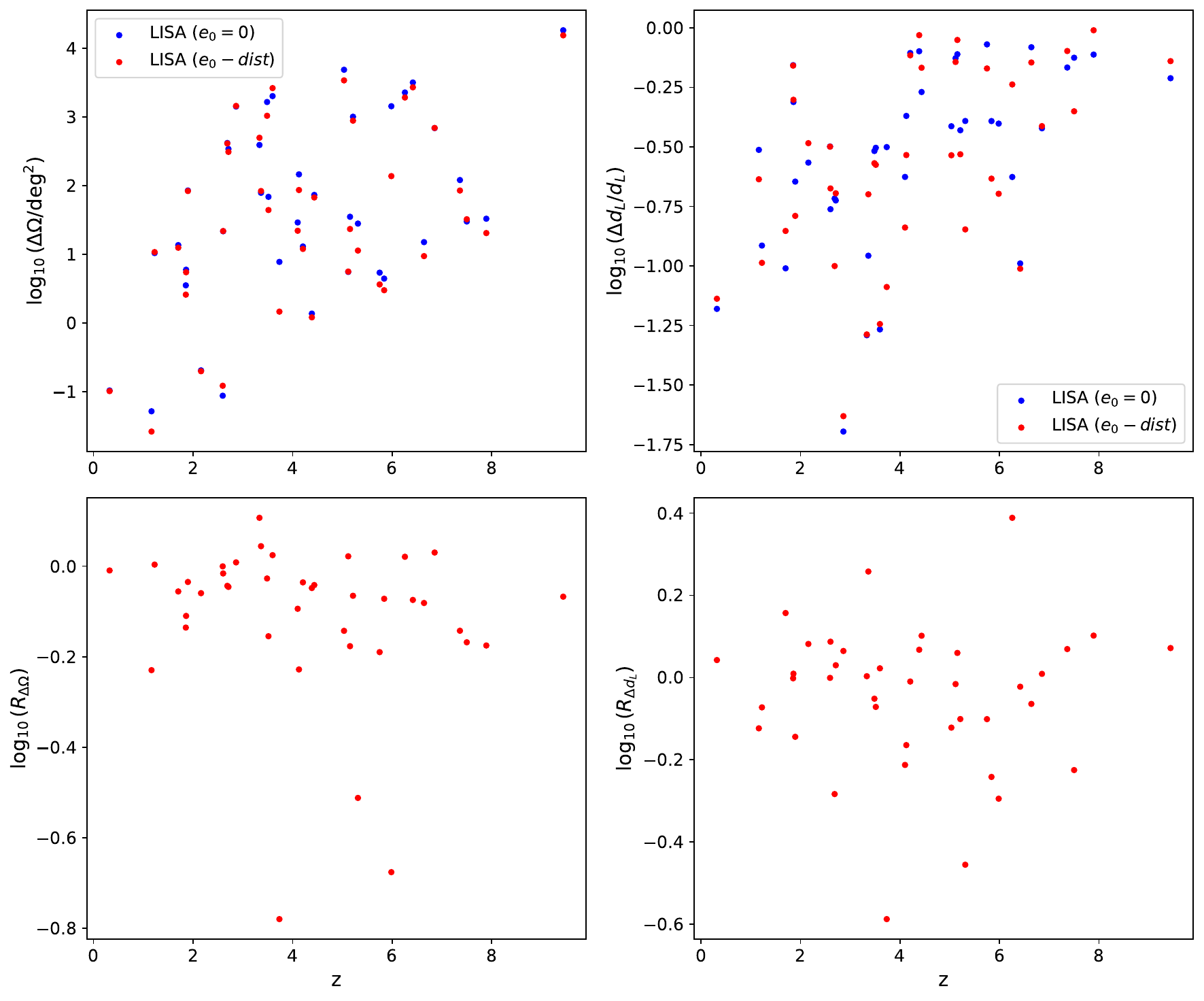}
    \caption{Same as Fig.~\ref{fig:Q3nod_location} in the main text, but for the Q3d model with orbital eccentricities drawn from the distribution shown in Fig.~\ref{fig:eccdis}.}
    \label{fig:Q3delocation}
\end{figure*}

\begin{figure*}[htbp]
    \centering
    \includegraphics[width=0.4\textwidth]{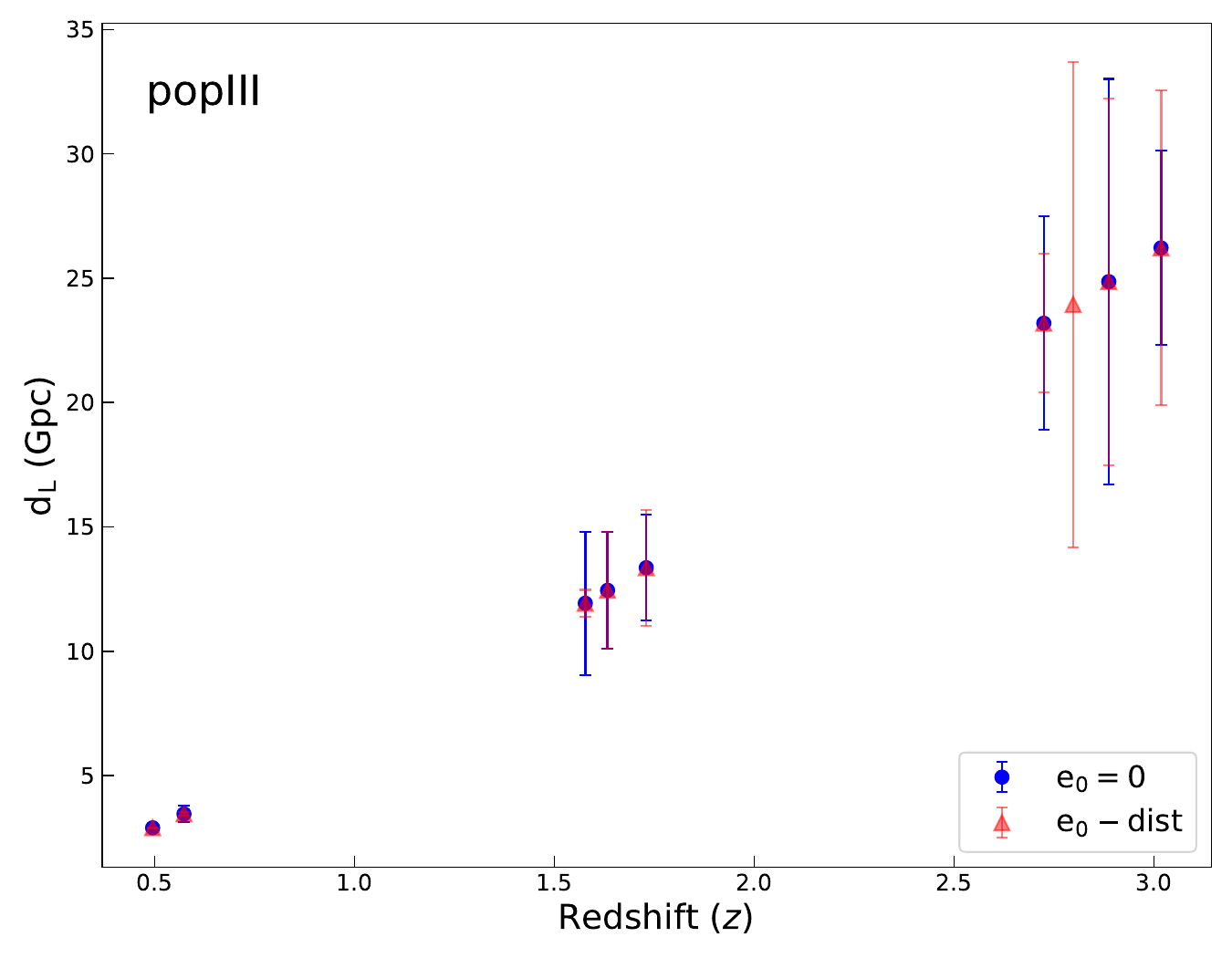}
    \includegraphics[width=0.4\textwidth]{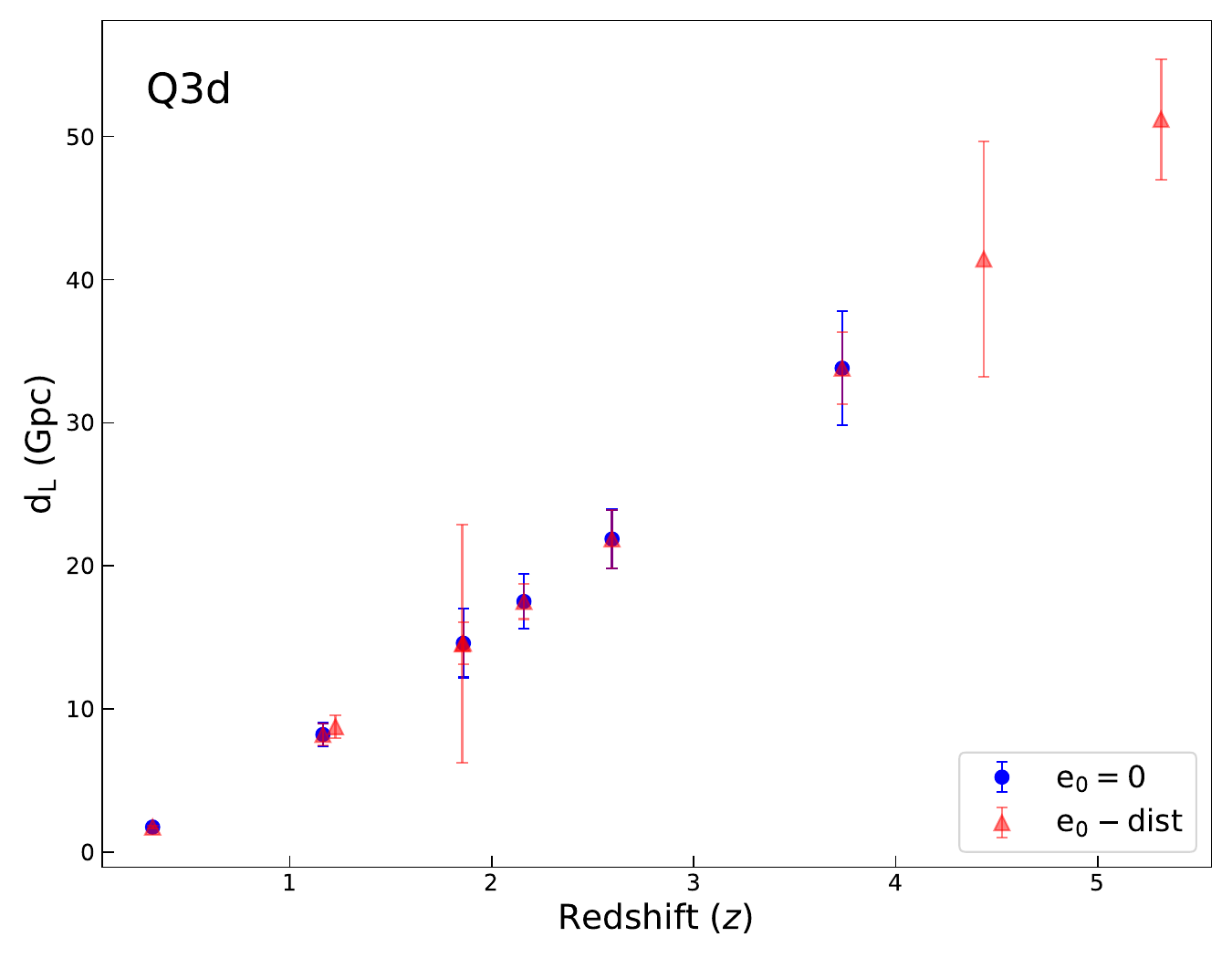}
    \includegraphics[width=0.4\textwidth]{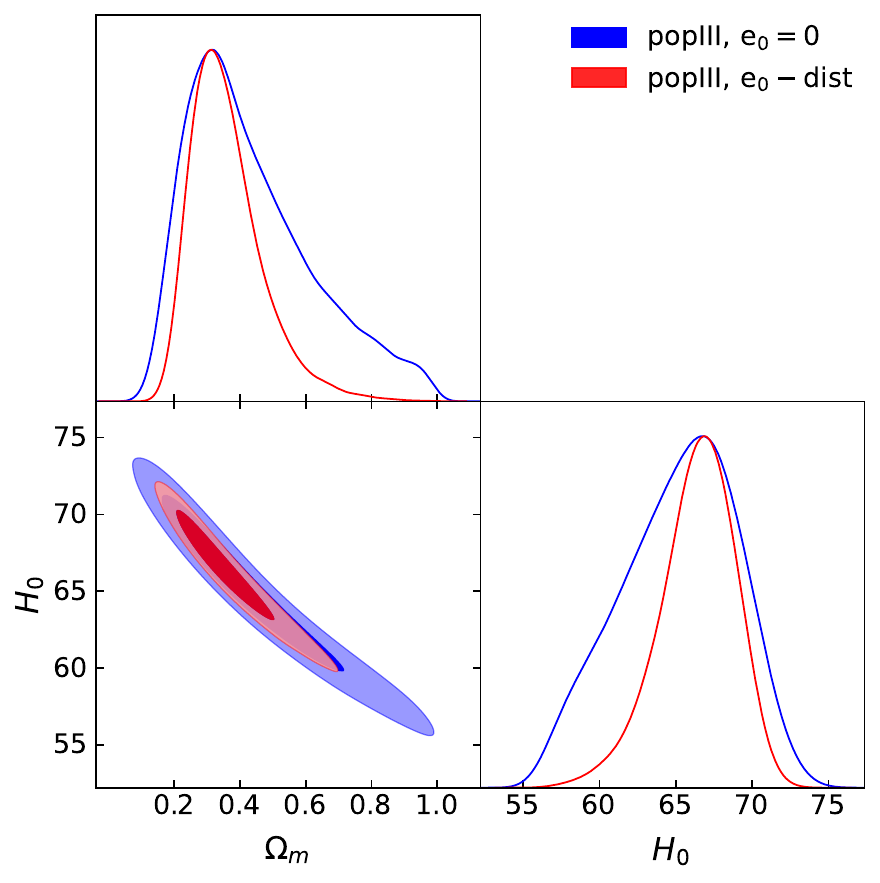}
    \includegraphics[width=0.4\textwidth]{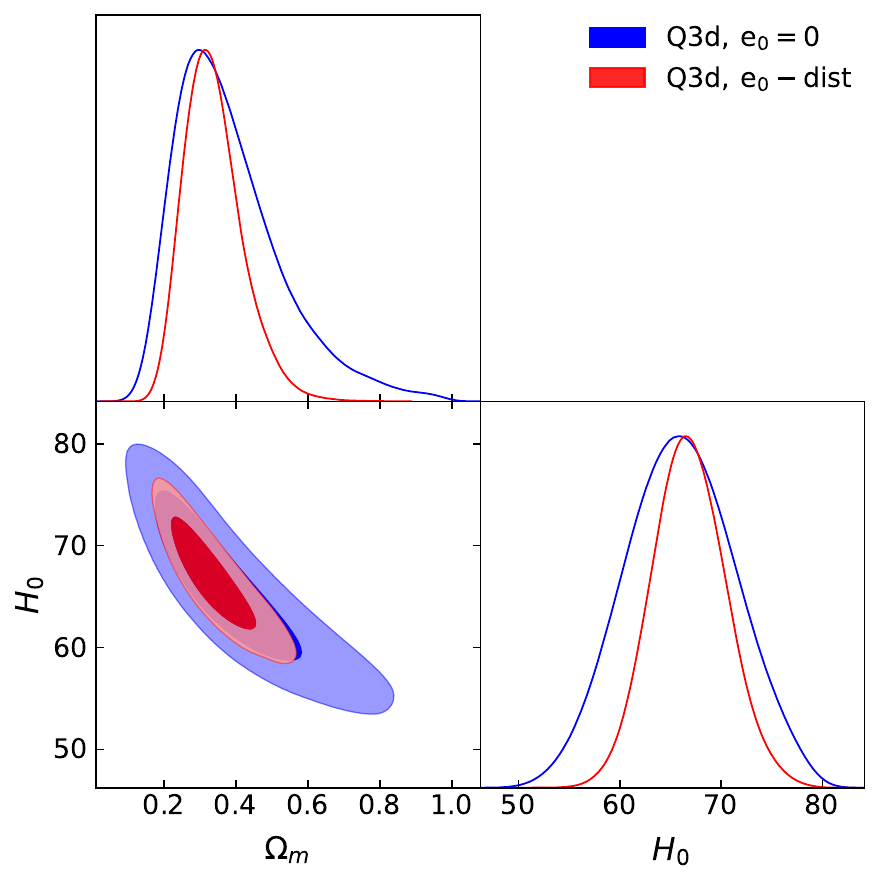}
    \caption{Top: Hubble diagrams of eccentric bright siren candidates observed by LISA over five years for the PopIII and Q3d models. Bottom: constraints on the Hubble constant $H_0$ and matter density parameter $\Omega_m$ in the $\Lambda$CDM model.}
    \label{fig:sts_e_cosmo}
\end{figure*}

\begin{figure}[htbp]
    \centering
    \includegraphics[width=0.45\textwidth]{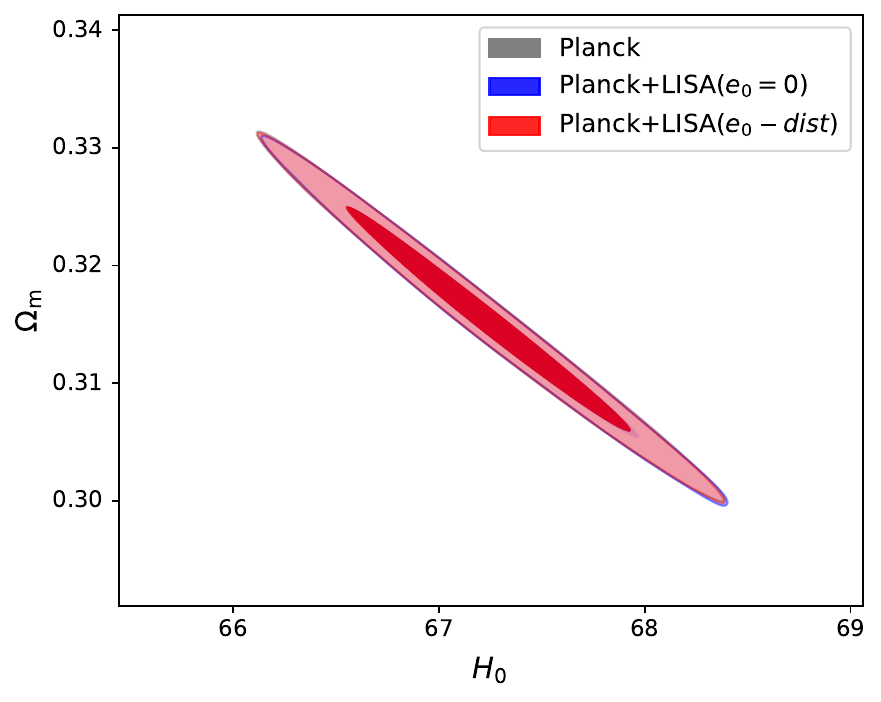}
    \caption{Joint constraints on $H_0$ and $\Omega_m$ in the $\Lambda$CDM model from CMB data and bright siren candidates in the Q3d model with orbital eccentricities drawn from the adopted distribution.}
    \label{fig:LCDM_Q3d_e}
\end{figure}

\begin{figure}[htbp]
    \centering
    \includegraphics[width=0.45\textwidth]{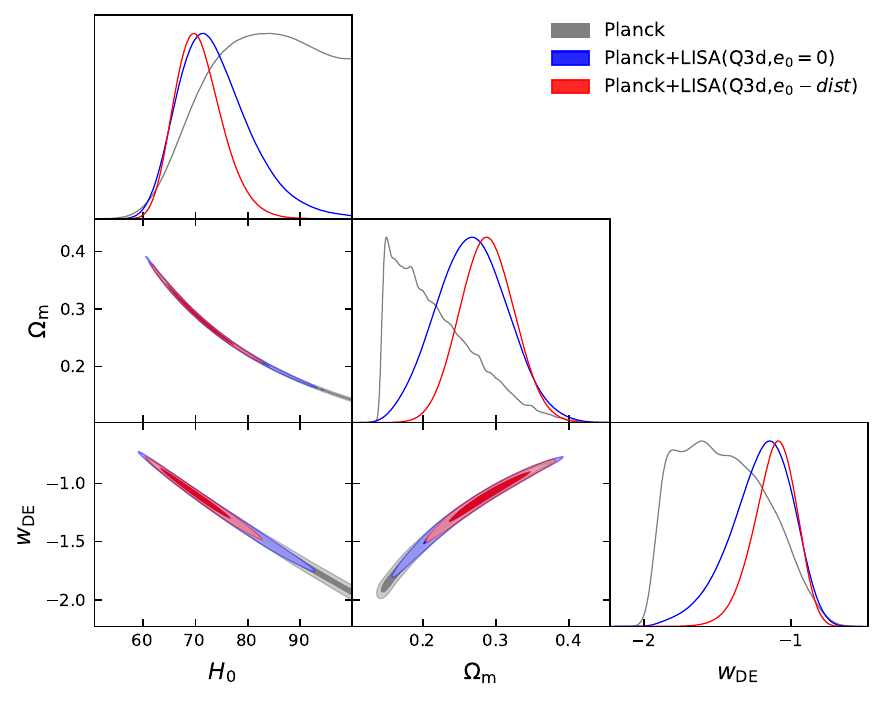}
    \caption{Joint constraints on $H_0$, $\Omega_m$, and $w$ in the $w$CDM model from CMB data and bright siren candidates in the Q3d model with orbital eccentricities drawn from the adopted distribution.}
    \label{fig:wCDM_Q3d_e}
\end{figure}

\begin{figure}[htbp]
    \centering
    \includegraphics[width=0.45\textwidth]{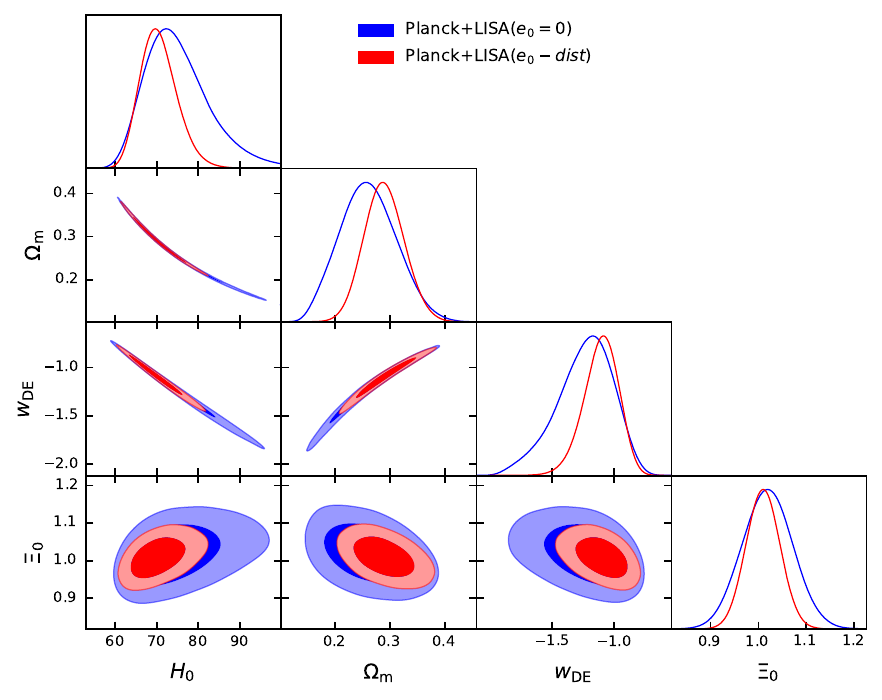}
    \caption{Joint constraints on $H_0$, $\Omega_m$, $w$ and $\Xi$ in the MG model from CMB data and bright siren candidates in the Q3d model with orbital eccentricities drawn from the adopted distribution.}
    \label{fig:MG_Q3d_e}
\end{figure}

\begin{acknowledgments}
T.Y. is supported by the National Natural Science Foundation of China Grant No. 12575063, and in part by ``the Special Funds for the Double First-Class Development of Wuhan University'' under reference No. 2025-1302-010. X.-H.D. is supported by the National Natural Science Foundation of China (Grant No. 12573017). K.L. is supported by the National Key R\&D Program of China (No. 2024YFC2207400). The numerical calculations in this paper have been done on the supercomputing system in the Supercomputing Center of Wuhan University.
\end{acknowledgments}

\section*{data availability}
The data that support the findings of this article are not publicly available. The data are available from the authors upon reasonable request.

\appendix
\section{QUASAR BOLOMETRIC LUMINOSITY \label{app:A}}
\setcounter{equation}{0} 
\renewcommand{\theequation}{A\arabic{equation}}
In this appendix, we describe how we estimate the quasar bolometric luminosity and convert it into an apparent magnitude. The bolometric luminosity $L_{\mathrm{bol}}$ is related to the absolute magnitude $M$ through~\cite{Zombeck_2006}
\begin{equation}
L_{\mathrm{bol}}=3.02\times 10^{35-\frac{2}{5}M}\,\rm erg/sec,
\end{equation}
while the relation between the apparent magnitude $m$ and the absolute magnitude $M$ is given by
\begin{equation}
m=BC+M-5+5 \log_{10}(\frac{d_L}{\rm pc}),
\end{equation}
where $BC$ is the bolometric correction and $d_L$ is the luminosity distance of the source. Therefore, in order to determine whether a flare associated with a massive black hole binary merger can be detected, we first compute its bolometric luminosity $L_{\rm bol}$ and then convert it into the apparent magnitude $m$.

The bolometric luminosity is assumed to be powered by accretion onto the massive black hole and can be expressed as
\begin{equation}
L_{\rm bol} = \dot{M}_{\rm bh,QSO} c^2 ,
\end{equation}
where $\dot{M}_{\rm bh,QSO}$ is the accretion rate during the quasar phase. Following the semianalytical prescription adopted in previous works, the accretion rate is limited by both the available gas reservoir and the Eddington rate,
\begin{equation}
\dot{M}_{\rm bh,QSO} = \min\!\left( \frac{M_{\rm res}}{t_\nu}, \, A_{\rm Edd}\, \dot{M}_{\rm Edd} \right) .
\end{equation}
Here $M_{\rm res}$ denotes the mass of gas reservoir feeding the MBH, while $t_\nu$ is the characteristic viscous timescale over which accretion proceeds. This timescale is approximated as~\cite{Barausse:2012fy,Sesana:2014bea,Antonini:2015cqa,Antonini:2015sza}
\begin{equation}
t_\nu \approx {\rm Re} \, t_{\rm dyn} ,
\end{equation}
where ${\rm Re} \approx 10^3$ is the critical Reynolds number marking the onset of turbulence, and $t_{\rm dyn} = \frac{G M_{\rm bh}}{\sigma^3}$ is the dynamical timescale evaluated at the black hole sphere of influence, with $M_{\rm bh}$ and $\sigma$ being the black hole mass and the stellar velocity dispersion, respectively. The Eddington mass accretion rate is defined as
\begin{equation}
\dot{M}_{\rm Edd} = \frac{L_{\rm Edd}}{\eta(a_{\rm bh}) c^2} ,
\end{equation}
where $L_{\rm Edd} = 1.26 \times 10^{38} \left( \frac{M_{\rm bh}}{M_\odot} \right) \, {\rm erg \, s^{-1}}$ is the Eddington luminosity and $\eta(a_{\rm bh})$ is the radiative efficiency, which depends on the dimensionless black hole spin parameter $a_{\rm bh}$. In this work, we adopt the free parameter $A_{\rm Edd} = 1$ for heavy-seed models and $A_{\rm Edd} \approx 2.2$ for light-seed models, following previous studies~\cite{Madau:2014pta}, in order to allow for mildly super-Eddington accretion.

Given the values of $M_{\rm bh}$, $a_{\rm bh}$, and $M_{\rm res}$ provided by the semianalytical galaxy formation model, we are able to compute $L_{\rm bol}$ and subsequently derive the apparent magnitude $m$. This procedure allows us to assess whether the electromagnetic counterpart associated with a given gravitational-wave event is detectable.

\section{PARAMETER ESTIMATION RESULTS OF POPIII AND Q3D MODEL CATALOGS FOR LISA \label{app:B}}
\setcounter{equation}{0}
\renewcommand{\theequation}{B\arabic{equation}}

In this appendix, we present additional results for the PopIII and Q3d population models observed by LISA, with particular emphasis on the impact of orbital eccentricity on source localization and distance inference at the catalog level.

Figure~\ref{fig:pop3location} shows the sky localization and luminosity distance distributions for the PopIII model. From the representative events discussed in Sec.~\ref{sec:catalog}, the improvement induced by eccentricity appears modest. However, a more detailed investigation of the full mock catalog reveals a non-negligible subset of low-redshift events for which the inclusion of eccentricity leads to a substantial enhancement in parameter estimation. In particular, for a fraction of PopIII events, the sky localization accuracy can be improved by up to $\sim 0.65$ orders of magnitude in the most favorable cases, reaching a level comparable to that obtained in the Q3nod and Q3d models. Similarly, the luminosity distance uncertainty can be reduced by up to $\sim 1$ order of magnitude for the best-constrained events. This behavior can be attributed to the presence of relatively more massive binaries within the PopIII catalog, for which higher harmonics induced by eccentricity enter the LISA band with sufficient SNR to significantly break parameter degeneracies. This effect provides a natural explanation for the fact that eccentricity also leads to a measurable improvement in cosmological parameter estimation for the PopIII catalog, despite the weaker enhancement observed in typical individual events. 

Figure~\ref{fig:Q3dlocation} displays the corresponding results for the Q3d model. The impact of eccentricity on parameter estimation is broadly consistent with that found for the Q3nod model. However, due to the inclusion of time delays between galaxy mergers and massive black hole coalescences, the Q3d model predicts a significantly smaller number of detectable events. As discussed in the Sec.~\ref{sec:cosmo}, when the number of available standard sirens is limited, the inclusion of eccentricity becomes even more important. In such a regime, the improved parameter estimation enabled by eccentric waveforms for individual events plays a crucial role in enhancing the overall precision of cosmological inference. Consequently, although the total event rate in the Q3d model is reduced, the relative gain from incorporating eccentricity is more pronounced and astrophysically more significant.

\section{CATALOGS OF LISA GWs WITH ECCENTRICITY DISTRIBUTIONS \label{app:C}}

In this appendix, we investigate the impact of a realistic orbital eccentricity distribution on the simulation of LISA catalogs and hence parameter estimation and cosmological inference. The analysis presented here complements the results in the main text and serves as a robustness check of our conclusions when eccentricity is drawn from astrophysically motivated distributions.

We adopt the eccentricity distributions presented in Ref.~\cite{Bonetti:2018tpf}, where a semianalytical model is employed to follow the cevolution of MBHBs and their host galaxies. That work considers three distinct evolutionary channels and two black hole seeding scenarios with different characteristic masses, which is consistent with the model used in our work. 

In Fig.~8 of Ref.~\cite{Bonetti:2018tpf}, the authors report the orbital eccentricity distributions evaluated at the ISCO for coalescing MBHBs with ${\rm SNR} > 8$. Using the relative detection rates of different channels as weights, we combine these distributions to construct an effective eccentricity distribution for the light-seed (PopIII) and heavy-seed (Q3d) models. The Q3nod model is not considered here, since the eccentricity primarily arises from time delay effects associated with environmental hardening, which are not included in the Q3nod scenario. 

To map the eccentricity from the ISCO to the LISA reference frequency $f_0 = 10^{-4}\,{\rm Hz}$, we employ the analytical relations between orbital eccentricity and frequency given in Eqs.~(2.4) and (2.5) of Ref.~\cite{Yunes:2009yz}. The resulting eccentricity distributions at $f_0 = 10^{-4}\,{\rm Hz}$ are shown in Fig.~\ref{fig:eccdis}.

Based on the eccentricity distributions shown above, we randomly sample orbital eccentricities when constructing the mock catalogs described in Sec.~\ref{sec:catalog}. We then select the corresponding bright siren events used in the cosmological analysis of Sec.~\ref{sec:cosmo}.
The resulting improvements in sky localization and luminosity distance inference are illustrated in Figs.~\ref{fig:pop3elocation} and \ref{fig:Q3delocation} for the PopIII and Q3d models, respectively. Hereafter, $e_0$-dist denotes the case in which orbital eccentricities are sampled from the distribution shown in Fig.~\ref{fig:eccdis}. Compared to the circular case and the $e_0 = 0.4$ case, the level of improvement lies between the two.

Using the same methodology as in Secs.~\ref{sec:catalog} and \ref{sec:cosmo}, we construct bright siren candidate catalogs for both population models and derive cosmological constraints under the $\Lambda$CDM model. The corresponding Hubble diagrams and posterior distributions obtained from bright sirens are shown in Fig.~\ref{fig:sts_e_cosmo}. Compared to the circular case, the inclusion of eccentricity leads to a noticeable tightening of the constraints, particularly for the Q3d model, for which the number of detected events is smaller. 

To further assess the robustness of our results, we combine these bright siren candidates catalogs with CMB data, following the same method as described in the main text.
For the $\Lambda$CDM model, the joint constraints from standard sirens and CMB data are shown in Fig.~\ref{fig:LCDM_Q3d_e}. Consistent with the results presented in the main text, since the constraints are already dominated by CMB data, the additional improvement from orbital eccentricity is limited.

We then extend the analysis to the $w$CDM model by allowing the dark energy equation-of-state parameter $w$ to vary. The resulting joint constraints are shown in Fig.~\ref{fig:wCDM_Q3d_e}. Even in this extended parameter space, eccentric bright sirens provide complementary information to the CMB, leading to improved constraints on $H_0$ and partially breaking degeneracies between $w$ and $\Omega_m$.

Finally, we consider MG scenarios using the same parametrization adopted in the main text. Consistent with the results presented there, the inclusion of eccentricity under a realistic eccentricity distribution also leads to a noticeable improvement in the parameter constraints. The results are shown in Fig.~\ref{fig:MG_Q3d_e}.

Overall, these results confirm that eccentric standard sirens provide robust and complementary cosmological information, both on their own and in combination with CMB observations, and that their impact becomes increasingly important when the number of detected events is limited.

\bibliographystyle{apsrev4-1}
\bibliography{ref}

\end{document}